# FAINT BLUE GALAXIES

*Richard S. Ellis*

Institute of Astronomy, University of Cambridge,
Madingley Road, Cambridge CB3 0HA, England
e-mail: rse@ast.cam.ac.uk

ABSTRACT: The physical properties of the faint blue galaxy population are reviewed in the context of observational progress made via deep spectroscopic surveys and Hubble Space Telescope imaging of field galaxies at various limits and theoretical models for the integrated star formation history of the Universe. Notwithstanding uncertainties in the properties of the local population of galaxies, convincing evidence has emerged from several independent studies for a rapid decline in the volume-averaged star-formation rate of field galaxies since a redshift $z \approx 1$. Together with the small angular sizes and modest mean redshift of the faintest detectable sources, these results can be understood in hierarchical models where the bulk of the star formation occurred at redshifts between $z \approx 1$-2. The physical processes responsible for the subsequent demise of the faint blue galaxy population remain unclear. Considerable progress will be possible when the evolutionary trends can be monitored in the context of independent physical parameters such as the underlying galactic mass.

KEY WORDS: distant field galaxies, evolution, galaxy formation, cosmology

## 1. INTRODUCTION

The nature of the faintest galaxies detectable with our large telescopes has a long and intriguing history and remains one of astronomy's grand questions. The quest to measure galaxy counts faint enough to verify the cosmological principle and to constrain world models motivated the construction of a series of ever larger telescopes and this, in turn, led to a renaissance of observational cosmology that has progressed apace during the twentieth century (Hubble 1926, 1936). The interest in the deep Universe as probed by faint galaxy statistics and the puzzling results obtained with modern instrumentation have been major driving forces in extragalactic astronomy through the 1980s and 1990s. The most recent chapter in this observational story is the Hubble Deep Field (HDF) project (Williams et al 1996) - an unprecedented long exposure of an area of undistinguished sky observed with the Hubble Space Telescope (HST) undertaken with the purpose of extending the known limits of the faint universe as delineated by field galaxies.

Although the original motivation for what has always been difficult observational work was an attempt to quantify the cosmological world model, the bulk of this review is concerned with the study of faint galaxies as a way of probing their evolutionary history. The evolutionary possibilities were, of course, recognized by the early pioneers (Hubble 1936, Humason et al 1956). However, galaxy evolution was surprisingly slow to emerge as the dominant observational motivation (see Sandage 1995 for a historical account). Even in the 1970s much of the relevant time on the Palomar 200-inch telescope was aimed at estimating cosmological parameters by using galaxies as tracers of the cosmic deceleration (Tammann 1984). Evolution was considered primarily as a necessary 'correction' to apply in the grander quest for the nature of the world model; the term 'evolutionary correction' remains (c.f. Poggianti 1997). Galaxy evolution only rose to prominence when quantitative predictions for the star formation histories of normal galaxies became available (Sandage 1961, Tinsley 1972). An equally important theoretical motivation for studying evolution was the development of the statistical machinery to understand how physical structures form in the expanding Universe (see e.g. Peebles 1971). Both propelled observers to construct catalogs of faint field galaxies and to consider comparing their statistical properties with theoretical predictions.

An appropriate birthplace for the modern era of faint field galaxy studies was the symposium 'Evolution of Stellar Populations' held at Yale University (Larson & Tinsley 1977). The theoretical ingredients were largely in place in time for the first deep optical surveys made possible by a new generation of wide-field 4-m telescope prime focus cameras. In an era in which the photographic plate is so often disregarded, it is salutary to realise that many of the basic observational results that generated the discussion that follows were first determined from photographic surveys exploited by the new generation of computer-controlled measuring machines (Kron 1978, Peterson et al 1979, Tyson & Jarvis 1979, Koo 1981).

The question of interpreting faint galaxy data in the context of the evolutionary history of field galaxies was comprehensively reviewed by Koo and Kron (1992). Their article summarised the developments from the early photographic results through to deeper charge-coupled derive (CCD) galaxy counts and, most importantly, provided a critical account of the interpretation of the first deep redshift surveys. As they stressed, the addition of spectroscopic data provided a valuable first glimpse at the distributions in redshift, luminosities and star formation characteristics of representative populations of field galaxies a few Gyr ago. Several puzzling features emerged that have collectively been referred to as the faint blue galaxy problem (Kron 1978). In its simplest manifestation, this is that an apparent excess of faint blue galaxies is seen in the source counts over the number expected on the basis of local galaxy properties. A more specific version of the problem that attracted much attention followed the results of the first faint redshift surveys (Broadhurst et al 1988, Colless et al 1990). The count-redshift data from these surveys did not solve the number problem by revealing a redshift range (at either low or high redshift) where this additional population could logically be placed. Relatively complex evolutionary hypotheses were then proposed to reconcile these results, including luminosity dependent evolution, galaxy merging and the existence of a new population of source present at modest redshift but, mysteriously, absent locally. Koo & Kron (1992) reviewed the material at hand but considered many of these hypotheses to be premature. They stressed the relatively poor knowledge of the local galaxy population and urged a more cautious approach. Meanwhile, however, the fundamental question remains as to how to account for the high surface density of faint blue sources seen to limits well beyond those of the spectroscopic surveys.

Although it is only 5 years since Koo and Kron's review, it is timely to address the question anew for several reasons. First, there has been an explosion of interest in the subject as evidenced by a number of diagnostics including large numbers of articles and the statistics of telescope time application. Clearly, in many minds, the controversy remains. Second, considerable observational progress has been made through larger, more comprehensive, ground-based redshift surveys including those from the first generation of 8- to 10-m telescopes. The redshift boundary for statistically complete field surveys has receded from $z=0.7$ (Colless et al 1990, 1993) to 1.6 (Cowie et al 1996). Third, significant new data bearing on the question has arrived from the refurbished HST including the Medium Deep Survey (MDS) (Griffiths et al 1994, Windhorst et al 1996) and the HDF (Williams et al 1996). Reliable morphologies and sizes of faint field galaxies have become available for the first time, providing a surge of new data similar to that provided by the first redshift surveys discussed by Koo and Kron.

The major questions addressed in this review are as follows: what are the faint blue galaxies seen in the deep optical galaxy counts and what role do they play in the evolution and formation of normal field galaxies; is there convincing evidence for recent evolution of the forms proposed and, if so, what are the physical processes involved? *Faint* in this context is defined, for convenience, to be at or beyond the limiting magnitude of the Schmidt sky surveys (i.e. V>21). *Faint* need not necessarily imply *distant* although a major motivation is the hope of learning something fundamental about distant young galaxies. The adjective *blue* is not applied rigorously in most of the articles in this area and the term primarily reflects the significance of the excess population when observed at optical wavebands sensitive to changes in the short-term star formation rate. *Field* implies selection without regard to the local environment (Koo & Kron 1992). Normally this is in the context of systematic surveys of randomly-chosen areas which span large cosmic volumes. The distinction between field and cluster observations is straightforward to apply but the physical significance of the different results remains unclear. The 1977 Yale symposium also saw the first quantitative evidence for an increase with redshift in the blue galaxy population in rich clusters (Butcher & Oemler 1978). Recent work (Couch et al 1994, Dressler et al 1994) interprets this evolution as being produced via a changing star formation rate in certain types of cluster galaxies, presumably as a result of environmental processes. However, until the physical cause of these activities are better understood, a connection between field and cluster evolutionary processes should not be ruled out.

Two points should be made about the scope of the review and its intended audience. First, a quantitative review of the evolution of galaxies would require a critical discussion of many related issues such as the stellar and dynamic history of nearby galaxies, indirect probes of the high redshift universe such as QSO absorption line statistics, primeval galaxy searches, the growth of large scale structure, uncertainties in stellar evolution and the reliability of evolutionary modelling used to interpret the wealth of data now available. All of these are active areas that impact heavily on galaxy evolution and merit reviews of their own. The strategy here is to focus solely on the faint blue galaxy question, drawing on additional evidence from these fields where appropriate. Secondly, the subject is developing rapidly with many new observational claims, some of which appear to contradict one another. A detailed resolution of these issues is usually highly technical and in many cases not yet possible. The review is therefore primarily aimed at a fresh graduate student entering the field rather than at the expert who is active in the subject. The hope is to bring out the key issues without getting overly bogged down in the numerical detail.

## 2. BASIC METHODOLOGIES

The primary observational material consists of a number of measurements for every galaxy in a survey that can be regarded as statistically complete. Completeness means that the results of one survey can be compared with those of other observers and with the predictions of contemporary models that take account of the various selection criteria. *Raw* measurements for each galaxy in a given survey typically include isophotal or pseudo-total magnitudes, aperture colors, spectroscopic redshifts and line strengths (such as [O II] 3727 A emission line fluxes or equivalent widths), dynamic line widths and, from HST, images, sizes and shapes. *Derived* pseudo-physical parameters for each galaxy include the luminosity in some rest-frame bandpass, the star formation rate as inferred from emission line characteristics or ultraviolet (UV) continuum flux, and some form of classification. Simple considerations (Struck-Marcell & Tinsley 1978) suggest that galaxies of similar Hubble types have had similar star formation histories providing the strong incentive to classify faint samples. This has been done variously via a morphological type from the HST image (Glazebrook et al 1995b, Driver et al 1995a,b), by examination of the color or spectral energy distribution (SED, Lilly et al 1995), or via cross-correlation of detailed spectral features against local spectroscopic data (Heyl et al 1997, Kennicutt 1992). Assuming completeness, *population statistics* can then be derived, such as the galaxy luminosity function and the volume averaged star formation rate, both as a function of redshift and galaxy type. These data can be used with other indicators, such as probes of the gas content, to infer the global history of star formation and metal production (Songaila et al 1990, Fall et al 1996, Madau et al 1996).

Before discussing how observers transform their raw measurements into population statistics, an important point of principle needs to be discussed. In the above methodology, the population statistics are used only to present an *empirical* description of galaxy evolution; no extensive modelling is usually involved although occasionally extrapolations are made into territories where no data exists. Recent examples of the empirical approach include redshift survey articles by Lilly et al (1996), Ellis et al (1996a) and Cowie et al (1996). An alternative approach, which we call the *ab initio* approach, starts from a cosmogonic theory. A particular initial power spectrum of density fluctuations is adopted and gas consumption timescales and morphological types are assigned to assemblies of dark and baryonic matter which grow hierarchically. Star formation histories for model galaxies are then used to predict observables directly in the context of particular surveys. The *ab initio* approach has its origins in the evolutionary predictions made by Tinsley (1980), Bruzual (1983), Guiderdoni & Rocca-Volmerange (1987) and Yoshii & Takahara (1988) on the basis of simple assumptions that did not rely on particular cosmogonic models. However, more elaborate calculations are now possible in the physical context of hierarchical dark matter cosmologies, such as in recent articles by Kauffmann et al (1994), Cole et al (1994a) and Baugh et al (1996).

Both the *empirical* and *ab initio* approaches have their place in observational cosmology. The *empirical* approach attempts to encapsulate the data in the simplest way and has the advantage of presenting results in a way that is not tied to any particular cosmogonic theory. However, it is not always clear how uncertainties in the raw data plane transform to those in the derived physical plane. The complexity of various selection effects is increasingly used in many areas of astronomy as an argument for the *ab initio* approach (cf simulations of the Lyman alpha forest discussed by Miralda-Escude et al 1996). Numerical simulations are now sufficiently sophisticated that remarkably realistic predictions can be made that take into account complexities such as merging, starbursts, feedback and ionization effects that affect model galaxies according to their physical situation rather than their observable quantities. In such cases, intuitive approaches are not possible. Ultimately the *ab initio* approach may become the preferred way to interpret data, but this will only occur when such models have strong predictive capabilities. The difficulty of relying entirely on the *ab initio* approach is that, at most, the current methodology reveals models that are only *consistent* with the observational datasets; one might argue this is the minimum required of *any* model! In no sense does agreement of a complex model with data imply that that particular model is correct.

## *2.1. Predictions in the Absence of Galaxy Evolution*

The *no evolution prediction* originated as a null hypothesis used to address the most basic question: Is evolution detected? Despite its frequent use, it clearly describes an unphysical situation. The passive evolution of stellar populations (defined as that arising from changes occurring in the absence of continued star formation) alone can lead to detectable luminosity and color changes over the past few Gyr (Tinsley 1972). However, the look-back time to a distant galaxy (which makes the entire subject of galaxy evolution observationally possible) is accompanied by redshift and related observational selection biases that seriously affect the empirical approach of monitoring evolution as a function of redshift in flux-limited catalogues of galaxies. For redshifts of z<1-2 the no evolution prediction has acted as a valuable standard baseline incorporating these biases from which the various evolutionary differences can be compared.

To predict the observables in the *empirical* approach, consider a local galaxy population with a luminosity function (LF) whose form is defined by the Schechter function (1976):

$$\phi(L)dL = \phi^* (L/L^*)^{-\alpha} \exp(-L/L^*) \, dL/L^* \quad [1]$$

Here $\Phi^*$ is a normalisation related to the number of luminous galaxies per unit volume, $L^*$ is a characteristic luminosity and $\alpha$ represents the shape of the function which determines the ratio of low luminosity to giant galaxies. The mean luminosity density for the Schechter function is:

$$\rho_L = \int L \, \phi(L) \, dL = \phi^* L^* \Gamma(\alpha + 2) \quad [2]$$

Assuming that the LF is independent of location and that galaxies are distributed homogeneously, the integrated source count to apparent magnitude m in the non-relativistic case is given by:

$$N(<m) \propto d^{*3}(m) \int dL \, \phi(L) \, (L/L^*)^{3/2} \propto \phi^* L^{*3/2} \, \Gamma(\alpha + 5/2) \quad [3]$$

where $d^*(m)$ is the limiting depth of the survey for a $L^*$ galaxy.

Equation [3] shows that the count normalization depends not only on $\phi^*$ but also on a good understanding of $L^*$ and $\alpha$. Recent determinations of the local field galaxy LF (Loveday et al 1992, Marzke et al 1994, Lin et al 1996, Zucca et al 1997) are somewhat discrepant in all three parameters. It is also important to determine the LF as a function of morphology or color because the visibility of the different galaxy types is affected by redshift bandpass effects. Color-based local LFs have been discussed by Metcalfe et al (1991) and morphological-based ones by King & Ellis (1985), Loveday et al (1992) and Marzke et al (1994).

The Schechter formula need not be adopted to make predictions. One could simply use the raw luminosity-redshift data for local surveys to predict the appearance of the faint Universe. Surprisingly, this has not been done. Indeed there is growing evidence that the simple Schechter expression fails to describe the full observational extent of the local field galaxy LF (Ferguson & McGaugh 1995, Zucca et al 1997). Studies of nearby clusters (Binggeli et al 1988, Bernstein et al 1995), where volume-limited samples can be constructed, also indicate significant departures from the Schechter form at luminosities $M_B > -16 + 5 \log h$[1]. An upturn at the faint end of the field galaxy LF could make a significant Euclidean contribution (i.e. $N \propto 10^{0.6m}$) to the faint number counts. Marzke et al's (1994) type-dependent LF suggests such a contribution would also be quite blue.

Given a LF for type j, $\phi(M_B, j)$, the type-dependent source counts $N(m,j)$ and redshift distributions $N(z, j)$ in flux-limited samples are calculable once the redshift visibility functions are known. In addition to the cosmological distance modulus, the type-dependent k-correction must be determined using the appropriate spectral energy distribution (SED), $f(\nu,j)$. Following Humason et al (1956):

$$k(z, j) = -2.5 \log(1+z) \int f(\nu,j) S(\nu) \Big/ \int f(\nu/1+z, j) S(\nu) \, d\nu \quad [4]$$

where $S(\nu)$ is the detector response function.

Progress in measuring the integrated SEDs of galaxies of different types over a wide wavelength range since the early studies of Pence (1976), Wells (1978) and Coleman et al (1980) used by King & Ellis (1985) because there has been no appropriate UV satellite with which *integrated*

---

[1] *h denotes Hubble's constant in units of 100 km s$^{-1}$ Mpc$^{-1}$*

*large aperture* SEDs can be determined. The lack of reliable k-corrections particularly affects the blue counts because, at high redshift, UV luminosities enter the visible spectral region. Kinney et al (1996) have obtained IUE-aperture spectrophotometry of 15 galaxies from $\lambda\lambda 1200$ A to 1μm (a technique attempted also by Ellis et al 1982) and claim that the small physical size of the areas sampled may nonetheless be useful if the SEDs span the range representative of the integrated galaxy light.

Ideally the k-correction should be averaged for galaxies of a given rest-frame color rather than morphology because the correlation between morphology and color is actually quite poor (Huchra 1997) and there are added complications for spheroidal galaxies arising from luminosity dependencies (Sandage & Visvanathan 1978). Recognising these difficulties, Bruzual & Charlot (1993) and Poggianti (1997) advocate determining the k-correction from model SEDs known to reproduce the integrated broad-band colors of real galaxies. A comprehensive analysis by Bershady (1995) indicates such a technique can match five-color data of a large sample of galaxies with z<0.3 with typical errors of only 0.04 mags. At higher redshifts, alternative routes to the k-correction have included interpolating SEDs from rest-frame colors (Lilly et al 1995) and matching spectroscopic data against local samples (Heyl et al 1997).

To what extent could k-correction uncertainties seriously affect our perceived view of the distant universe? The uncertainty in the optical k-correction for the bulk of the Hubble sequence is probably fairly small to z=1-1.5 (Figure 1a). However, this conclusion is based on optically selected samples. Although Bershady (1995) found a few galaxies with colors outside the range expected on the basis of his modeling, for high redshift work the test should be extended into the UV. Serious errors of interpretation at fairly modest redshifts could occur if there existed a population of galaxies whose UV-optical colors did not match the model SEDs. Using balloon-borne instrumentation, Donas et al (1995) present a large aperture UV (2000 A)-optical color distribution of a sample of galaxies *selected in the UV*; a significant proportion of this sample at all optical colors show UV excesses compared to the conventional range of SEDs (Figure 2). Redshifts are needed to derive luminosity densities for these galaxies in order to quantify their possible effect on the k-corrections in use. Such inconsistencies in our understanding of the UV continua of galaxies indicate much work remains to be done in this area as well as highlighting the continued need for survey facilities at these wavelengths.

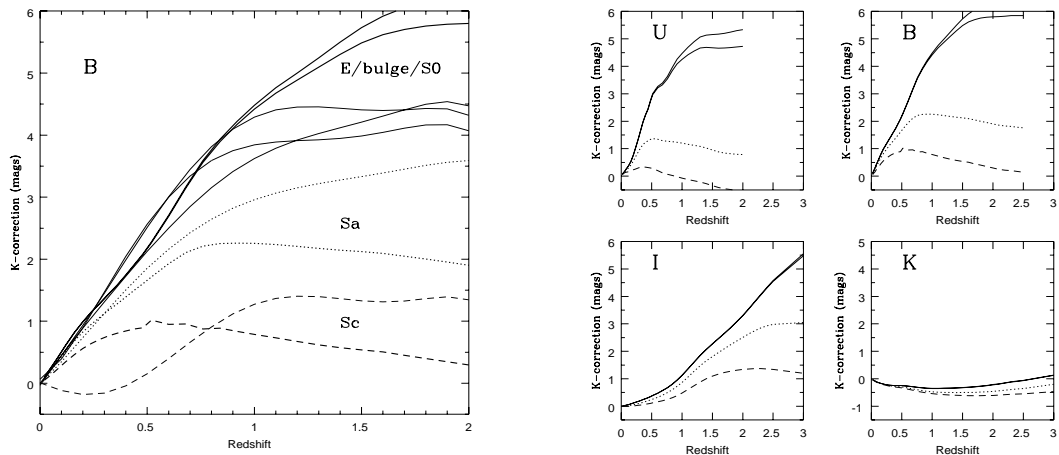

*Figure 1: (a) Type-dependent k-corrections for the B-band using two different approaches. Bold lines indicate corrections based on model SEDs that reproduce integrated broad-band colors (Poggianti 1997); non-bold lines indicate corrections derived from aperture spectrophotometry analysed in conjunction with color data (Kinney et al 1996). (b) The k-correction for ellipticals (solid lines), Sa (short-dash) and Sc (long-dash) galaxies for the UBIK photometric bands from the models of Poggianti (1997).*

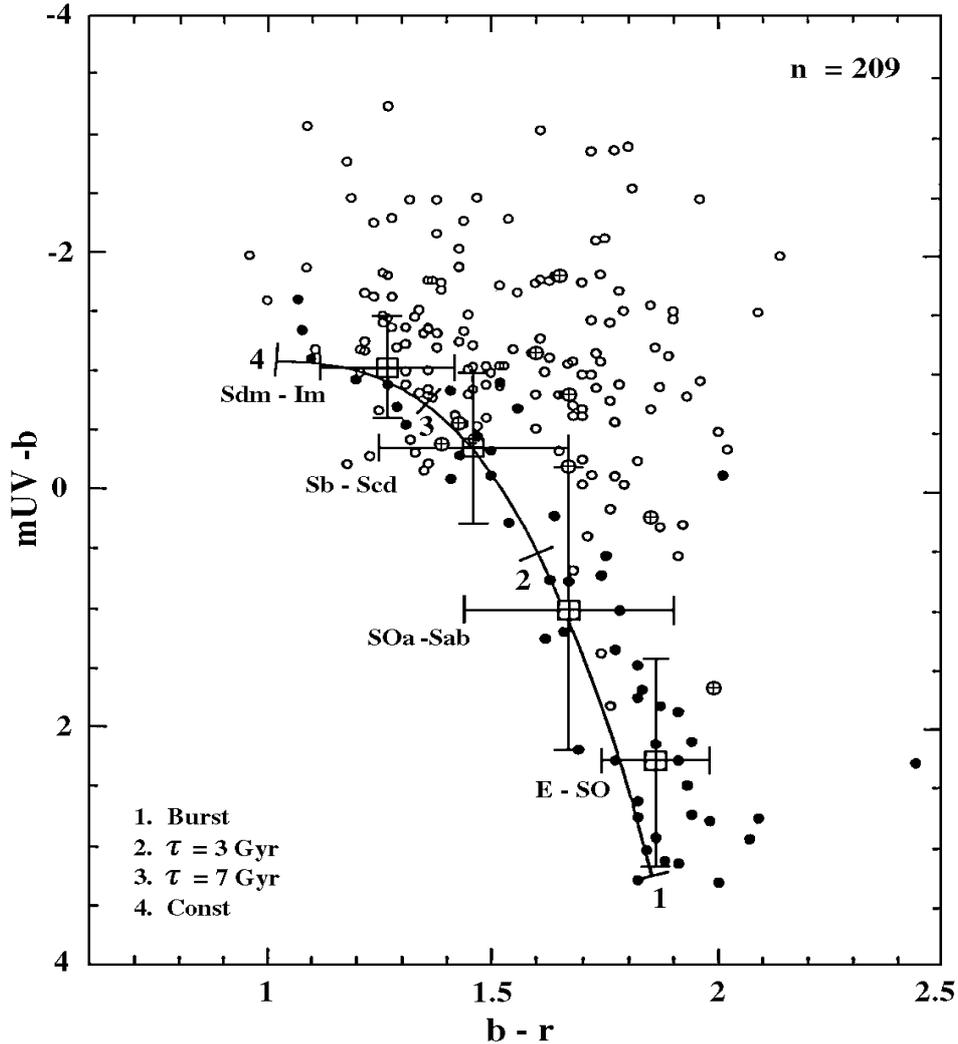

*Figure 2: UV (2000 A) - optical colors for a sample of UV-selected galaxies from the balloon-borne imaging study of Donas et al (1995). A significant proportion of the sample appears to have normal b-r colors but UV excesses of 1-2 magnitudes compared to standard models (solid curve; Bruzual & Charlot 1993) used to infer k-corrections for the field galaxy population.*

Inspection of the k-correction as a function of observed passband (Figure 1b) reveals the important point that the variation with type or optical color is considerably reduced in the near-infrared (Bershady 1995). There is also the benefit that, to high redshifts, the required SEDs are based on large aperture optical data. The smaller k-corrections reflect the dominant contribution of red giants at longer wavelengths in old stellar populations and has been used to justify a number of faint K-band surveys for evolution and cosmology (Gardner et al 1993, Songaila et al 1994, Glazebrook et al 1995c, Djorgovski et al 1995, Cowie et al 1996) on the grounds that their interpretation is less hampered by k-correction uncertainties. By implication, the observed mixture of types at faint limits is less distorted to star forming types than is the case in optical surveys. However, if the primary motivation of a faint survey is to learn about the star formation history of field galaxies, the gains of an infrared-selected survey are somewhat illusory. The insensitivity of the K-band light to young stars means that a deep infrared survey is primarily tracing the mass distribution of distant galaxies. When optical-infrared colors are introduced to aid the interpretation (Cowie et al 1996), the poorly-understood UV continua return to plague the analysis.

The cosmological sensitivity that originally motivated the topic of galaxy counts (Hubble & Tolman 1935, Hubble 1936, Sandage 1961) enters through the luminosity distance, $d_L(z)$ (Weinberg 1974, Peebles 1994) and the volume element $dV(z)$. The apparent magnitude of a faint galaxy of absolute magnitude M is determined by both the luminosity distance, $d_L$ (z, $\Omega_M$, $\Omega_\Lambda = \Lambda/3H_o^2$) in Mpc (Carroll et al 1992, eqn 25) and the type-dependent k-correction. Because the flux received from an apparent magnitude m scales as $10^{-0.4m}$, the contribution that the differential counts N(m) make to the extragalactic background light (EBL) is obtained via the following (Mattila et al 1991, Vaisanen 1996):

$$I_{EBL} \propto \int N(m)\, 10^{-0.4m}\, dm \qquad [2.9]$$

Given the local properties of galaxies, the joint distribution N(m,z,j) can be predicted and compared with faint datasets. As Koo & Kron (1992) discussed, the availability of redshift data enables far more sensitive tests than for photometric data alone. Some information on the redshift distribution of faint sources can be derived from multi-band colors. However, redshifts based on colors suffer from similar uncertainties to those discussed earlier for the k-correction. If local SEDs do not fully span those sampled in faint data, systematic errors may occur. The technique has a long history beginning with Baum (1962), Koo (1985) and Loh & Spillar (1986, who sought to constrain cosmological models via volume tests). In recent years, more sophisticated techniques have been introduced that rely on spectroscopic confirmation for some subset of the sample, on the use of magnitude information to break various degeneracies (Connolly et al 1995), or on strong discontinuities in the spectral energy distributions (Guhathakurta et al 1990, Steidel & Hamilton 1992, Madau et al 1996). For extremely faint galaxies, color-based redshifts are increasingly used as a way of locating information for sources beyond the spectroscopic limits (see Section 5).

## *2.2 Evolutionary Predictions*

The no evolution models defined earlier have been very useful as a baseline for comparing the predictions and observations made by different workers but, as the datasets available probe to higher redshift, evolutionary modeling has become increasingly important. These models take as ingredients the stellar evolutionary tracks (normally for a restricted metallicity range), the initial mass function and a gas consumption timescale adjusted to give the present range of colors across the Hubble sequence. Assuming each galaxy evolves as an isolated system, the rest-frame SED can be predicted at a given time and thus an evolutionary correction can be determined with respect to the no evolution equations above, or the predictions can be incorporated in *ab initio* models to generate simulated datasets. The cosmological model is a crucial, but often overlooked, variable in linking time and redshift. For $H_o$=70 and $\Lambda$=0, the redshift corresponding to a lookback time of, say, 7 Gyr, varies from z=1-3 depending on $\Omega$.

A discussion of the reliability of these evolutionary predictions and the differences between the various approaches is beyond the scope of this review. Physical difficulties arise from the degenerate effects of age and metallicity and the uncertainties of post-main sequence stellar evolution. Results for various models have been intercompared by Mazzei et al (1992), Bruzual & Charlot (1993) and, most recently for populations of the same input age and metallicity, by Charlot et al (1996). Recent efforts have concentrated on improving the stellar tracks and spectral libraries (Bruzual & Charlot 1993) and including the effects of chemical evolution (Arimoto et al 1992). An all-inclusive database of progress in this area is presented by Leitherer et al (1996).

For *single burst populations* presumed appropriate for early-type galaxies, Charlot et al discuss surprisingly large discrepancies in the predicted behaviour of such populations at times after 1 Gyr. The differences amount to at least 0.03 mag in rest-frame B-V, 0.13 mag in rest-frame V-K and a 25% dispersion in the V-band mass/light ratio. The large uncertainties in the predicted optical-infrared colors and visual luminosity evolution imply a significant age range (4-13 Gyr) that is permissible even for the simplest case of a passively-evolving red galaxy, emphasising the continuing need to compare these models with representative high redshift data as well as the need to improve our knowledge of post-main sequence stellar evolution.

For populations with *constant star formation*, both the evolutionary corrections and the discrepancies between the available models are less. This is because the same main sequence stellar types dominate the spectra at most times and their theoretical behaviour is considerably better understood. Unfortunately, precise predictions are required for *both* types of model galaxy (as well as the large range in between) because faint blue galaxies could be either passively-evolving systems seen at an early stage, systems of constant star formation or bursts of star formation imposed on a quiescent system. A comparison of the predicted evolutionary behaviour for two of these cases is shown in Figure 3. Such uncertainties represent a formidable obstacle to detailed modeling in the *ab initio* approach.

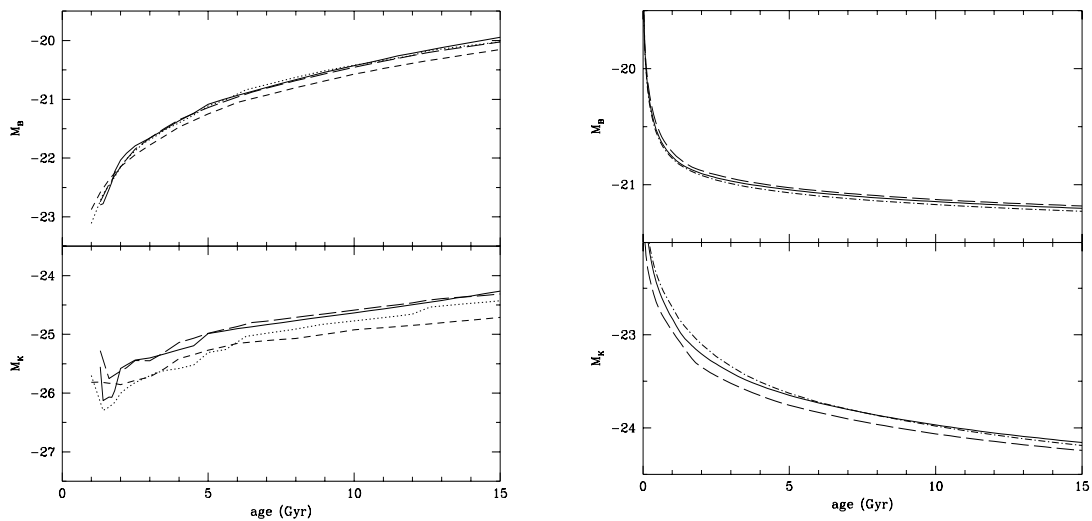

*Figure 3: Evolution of the rest-frame B and K band luminosity with time according to recent models by Bruzual & Charlot (1993, solid line), Bertelli et al (1994, dotted) and Worthey (1994, short-dash); long-dash represents a Bruzual & Charlot model based on alternative evolutionary stellar tracks (see Charlot et al 1996 for details). (a) Comparison for a passively-evolving galaxy following an instantaneous burst with solar metallicity (see Charlot et al 1996 for details); (b) comparison for a constant star formation model (see Charlot 1996 for details).*

## 2.3 Images from HST

Since the review by Koo & Kron (1992), multi-color HST images have become available for galaxies to the limits achievable with ground-based spectroscopy and beyond. The images are spectacular and provide morphological data for large numbers of distant galaxies, a growing number of which have redshifts. The potential for directly tracking the evolution of the Hubble sequence of types is clearly an exciting opportunity. However, the interpretation of HST data in this way has not been straightforward (cf discussion by Abraham et al 1996a,b). Although many workers have concluded that the faint field population is dominated by `irregular' and `merging' galaxies (Glazebrook et al 1995b, Driver et al 1995b), both the reliability of this result and a physical understanding of what it signifies requires careful study. When comparing distant images with possible local counterparts, the reduced signal to noise ratio, relative increase in background and cosmological surface brightness dimming tend to accentuate the visibility of high contrast features. Furthermore, bandpass shifting effects arising from the k-correction mean that low redshift optical images are being compared with high redshift UV images (O'Connell & Marcum 1996). An illustration of these effects is given in Figure 4 following the work described by Abraham et al (1996b). Such studies indicate that, although the broad classification system (E/Spiral/Irr) is reliable to z$\approx$1 for most large regular galaxies, any unfamiliarity in the images of higher z systems may be spurious.

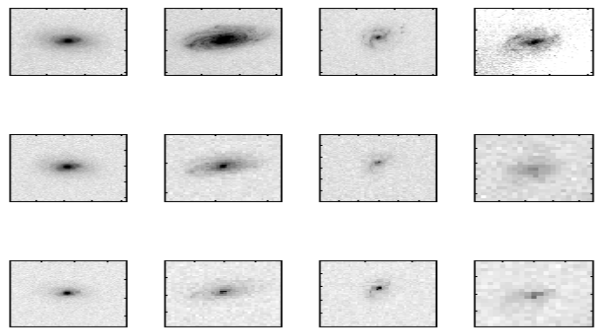

*Figure 4: Simulated appearance of NGC 4450 (Sab), 3953 (Sbc), 5669 (Scd) and 4242 (Sdm)) at various redshifts. With the exception of NGC 4242, which is sub-luminous, all galaxies have $L \approx L^*$. From top to bottom the panels show (a) the restframe B band images, (b) the appearance as viewed at z=0.75 in a three orbit F814W WFPC-2 image and (c) the appearance at z=1.5 in an exposure equivalent to the HDF. (Courtesy of Roberto Abraham).*

The concept of comparing HST images of galaxies *grouped by class* at different redshifts raises a very important problem. Most of the above discussion has centered on statistical results from surveys. By examining the *entire* field population accessible at each redshift, robust statements are made about the global changes in, for example, the galaxy LF and volume averaged star formation rates. Yet, to *physically understand* these changes in terms of the elaborate models now available we are encouraged to divide the samples into *subclasses* in the hope of determining the evolutionary behaviour as a function of type. This can only be done if there is some fundamental slowly-changing property of a galaxy that can act as a label. Clearly, both color and spectral class (cf Lilly et al 1995, Heyl et al 1997) could be transient properties affected by short-term changes in star-formation induced, for example, by merging or other processes. Likewise galaxy morphology, even if correctly assigned in HST images that sample a range of rest-frame wavelengths, may change with time. White (1996) discussed physical situations where gas-rich disk galaxies may merge to form gas-poor spheroidals only to accrete more gas and become spirals again. The possible migration of galaxies in and out of a *faint blue category* by some unidentified time-dependent process may be an important stumbling block to progress.

## 3. RECENT OBSERVATIONAL RESULTS

### 3.1 Number Magnitude Counts

Large format CCD and infrared-sensitive detectors have been used to extend ground-based number counts since Koo & Kron's (1992) review. The greater depth now available has confirmed a break in the count slope that occurs, respectively, at around B$\approx$25 (Lilly et al 1991, Metcalfe et al 1995b) and K$\approx$18 (Gardner et al 1993, Djorgovski et al 1995, Moustakas et al 1997). The slope, $\gamma$ = dlogN/dm, ranges from $\gamma_B$ = 0.47 to 0.30 and from $\gamma_K$ = 0.6 to 0.25, but the surface density at the B break is $\approx$30 times higher than that at the K break. If the two effects were manifestations of the same phenomenon, e.g. a decline in say the volume density beyond some redshift limit, then the mean B-K color should not change significantly across the break points. However, the change in slope is accompanied by a marked increase in the number of galaxies with colors B-K < 5, which explains the steeper $\gamma_B$ at fainter limits. Such blue galaxies were originally referred to as *flat spectrum* galaxies by Cowie et al (1989) because their SEDs approximate ones with f($\nu$) = constant.

The break in slope and, more importantly, the disparate behaviour between the B and K counts with respect to the no evolution predictions makes it unlikely that a major portion of the excess counts arises via a non-zero cosmological constant, $\Lambda$, as postulated by Fukugita et al (1990) and Yoshii & Peterson (1991) (see Carroll et al 1992 for a full discussion). That such a dramatic excess should be seen in B but not K could only be consistent with the hypothesis of a non-zero cosmological constant if the B>25 sources were significantly more distant than the K>19 ones, which seems unlikely (Section 5). Note again that the K-counts, by virtue of their insensitivity to the k-correction (Fig. 1b), remain an interesting cosmological probe although a satisfactory conclusion concerning $\Omega$ and $\Lambda$ will remain elusive so long as the evolutionary behaviour is poorly-determined (Djorgovski et al 1995).

The break in slope more probably signals a transition in galaxy properties at some redshift with the bulk of the fainter sources being drawn from intrinsically less luminous sources at similar redshifts, a conclusion favored by Lilly et al (1991), Gardner et al (1993) and Metcalfe et al (1995b). In this case, the abundance of fainter blue sources would suggest a high volume density of low mass galaxies and the slope of the B and K counts would directly reflect their relative contribution to the luminosity function at that time. Indeed, from equation [5] clearly the break indicates the apparent magnitude beyond which the contribution of galaxies to the extragalactic background begins to converge. Specifically, in the B band that samples the rest-frame UV at the appropriate redshifts, the location of the break defines, albeit qualitatively, that era in which galaxies contribute most in terms of short-term star formation and associated metal production (Songaila et al 1990, Lilly et al 1996, Section 6).

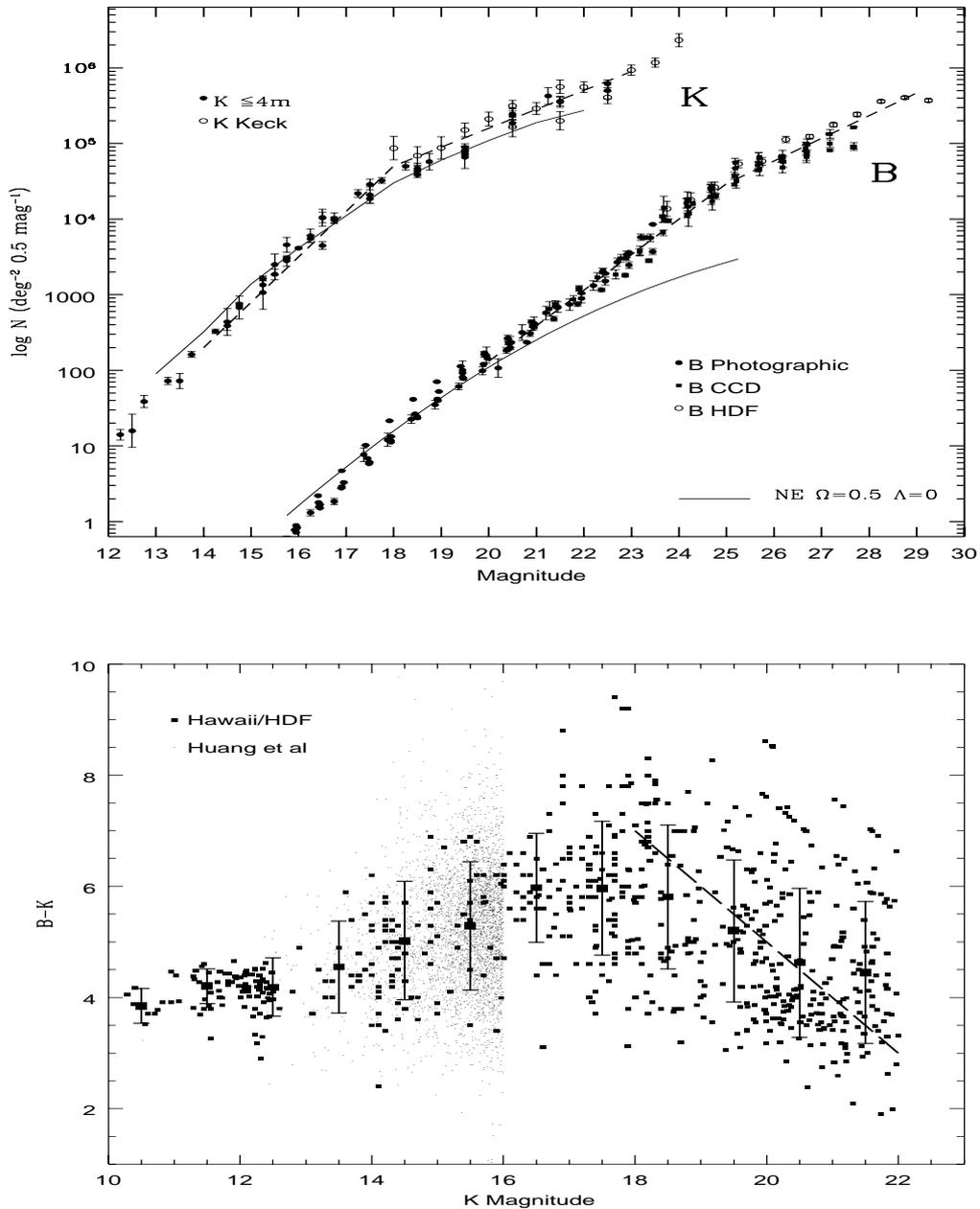

*Figure 5: (a) Differential galaxy number magnitude counts in the B and K passbands from the compilation of Metcalfe et al (1996) augmented with the Keck K counts of Moustakas et al (1997). The K counts have been offset by +1 dex for clarity. The two power law slopes (dashed lines) drawn have $\gamma$ (=dlog N/dm) = 0.47,0.30 around B=25 (Metcalfe et al 1995a) and 0.60,0.25 around K=18 (Gardner et al 1993). The solid curve indicates a no evolution prediction for an Einstein-de Sitter universe based on King & Ellis' (1985) k-corrections and luminosity functions from local surveys with a normalisation raised by 50% (see Section 4 for discussion). Beyond the limits drawn, the no evolution prediction ceases to be reliable. (b) B-K versus K for galaxies from the Hawaii bright survey (fine dots, Huang et al 1996), the Hawaii deep surveys (Cowie et al 1996 and references therein) and HDF (Cowie 1996). Points with error bars refer to mean colors for various magnitude slices. The dashed line indicates the current limit for optical spectroscopy.*

## *3.2 Faint Redshift Surveys*

Ground-based redshift surveys have not yet convincingly penetrated past the break point in the optical counts to demonstrate the suggestions above. The bulk of the published work has been concerned with tracking the evolution of the galaxy LF up to B=24, I=22 and K=18; these are the effective limits in reasonable integration times on 4-m class telescopes. Although redshifts are available beyond these limits from the first Keck exposures (Koo et al 1996, Illingworth et al 1996), with the exception of the Cowie et al (1996, 1997) surveys, these do not yet constitute a controlled magnitude-limited sample. A summary of the published (and where known, unpublished) faint field redshift data is given in Table 1. The Kitt Peak survey discussed provisionally by Koo & Kron in their 1992 review is to be published in a series of forthcoming papers commencing with Munn et al (1997).

**Table 1 - Recent Faint Galaxy Redshift Surveys**

| Reference | Survey[1] | Magnitude range | Redshifts |
|---|---|---|---|
| Lilly et al (1995) | CFRS | $17.5<I<22$ | 591 |
| Ellis et al (1996a) | Autofib/LDSS | $11.5<b_J<24$ | 1726 |
| Cowie et al (1996) | Keck LRIS | $14<K<20$ | 346 |
|  |  | ($I<23$ | 287) |
| Cowie et al (1997) | Keck LRIS | $B<24.5$ | 166 |
| Lin et al (1996b) | CNOC1 | $r<22$ | 389 |
| Koo et al (1996), Illingworth et al (1996) | Keck DEEP | $I<25$ | ≈115 |
| Munn et al (1997) | KPGRS | $R<20+$ | 739 |

The survey data suggest a mean redshift z ≈0.8 at the break point B=25, although no complete survey this faint yet exists. The surveys are still incomplete at some level beyond B≈23 and I≈21 and this can seriously affect the inferred redshift distribution. There is, for example, a significant difference between the faint B<24 LDSS-2 redshift distribution (Glazebrook et al 1995a) used by Ellis et al (1996a) and that presented by Cowie et al. (1996), arising presumably from a loss of higher redshift galaxies whose [O II] emission was redshifted out of the LDSS-2 spectroscopic window. Such incompleteness will remain a concern until infrared spectrographs are available that can systematically track [O II] and Hα emission to z≈2 (Ellis 1996b).

The most significant results from the redshift surveys are the redshift and type-dependent LFs. Furthermore, provided the spectroscopic surveys sample representative regions sufficiently faint to probe the apparent excess population, physical models that account for the counts to fainter apparent magnitude limits can also be tested. By combining data sampling a wide range of apparent magnitude, a range in luminosity is available for z<0.5 (for the B-selected survey of Ellis et al 1996a) and z<1 (for I- and K-selected surveys of Lilly et al 1995 and Cowie et al 1996). The LFs so derived indicate little change in the volume density of luminous galaxies to z=1 whereas less luminous galaxies appear to evolve more rapidly. Ellis et al (1996a) presented evidence for a steepening with redshift of the faint end slope of the overall LF consistent with a rising contribution of less luminous galaxies. The effect is particularly strong when parameterised according to [OII] emission line strength (Ellis et al 1996a) and UV-optical color (Cowie et al 1996). The bulk of the excess population to B=24 compared to the no evolution predictions (Figure 5a) may be largely due to this phenomenon (Glazebrook et al 1995a).

---

[1] *CFRS=Canada France Redshift Survey(CFHT); Autofib=multifiber spectrograph (AAT); LDSS=Low Dispersion Survey Spectrograph (AAT/WHT); CNOC1=Canadian Network for Observational Cosmology (field component of cluster survey, CFHT); KPGRS=Kitt Peak Galaxy Redshift Survey; DEEP=Deep Extragalactic Evolutionary Probe (Keck).*

Such *luminosity-dependent* evolution was originally proposed by Broadhurst et al (1988) to reconcile their redshift survey with photometric data at brighter limits (B<21.5). However, strong evolution at these limits, corresponding to z≈0.25, now seems less likely because of growing evidence for a higher normalisation of the local counts. The early B-selected surveys (Broadhurst et al 1988, Colless et al 1990, 1993) used the Durham-AAT Redshift Survey (DARS: Peterson et al 1986, Efstathiou et al 1988) and APM-Stromlo (Loveday et al 1992) datasets as local benchmarks. As Ellis et al (1996a) discuss, the low z LF derived from the fainter surveys indicates a higher normalisation. An upward revision, consistent with the galaxy counts at B≈19, would reduce the original dilemma presented by Broadhurst et al and Colless et al whereby excess galaxies are seen in the counts but within the same redshift range as expected in the no evolution case. The implication of this change may be that some local data is unrepresentative, possibly because of photometric difficulties at bright apparent magnitudes or because the survey regions used are deficient in galaxies for some reason. The effect is apparent also in the source counts (Maddox et al 1990).

Given the uncertainties in the normalisation of the local benchmark samples, Lilly et al (1995) argue that greater reliance should be placed on evolutionary trends determined internally from self-consistent datasets. The CFRS survey has the considerable advantage of probing 0.2<z<1 on the basis of a single well-defined photometric scale. The limited apparent magnitude range sampled offers less conclusive results on possible evolution in the *shape* of the LF (Figure 6a) but the luminosity-dependent trends are similar to those in the other surveys. Whether the shape is changing or the overall LF is brightening can be judged to a limited extent by examining the contrasting behaviour of the CFRS and Autofib/LDSS LFs in the redshift ranges where both have reasonable sample sizes (Figure 6b). The CFRS results do admit some luminosity evolution at the bright end whereas the changes in the Autofib/LDSS LF occur solely for the less luminous objects.

The underlying trend supported by all the major redshift surveys is a marked increase in the volume density of star-forming galaxies with redshift. That this evolution occurs primarily in sub-luminous galaxies is less clear. Lilly et al (1995) support the distinction in the evolutionary behaviour of galaxies redder and bluer than the rest-frame color of a typical Sbc spiral. Cowie et al (1996) conclude the evolution proceeds according to a `down-slicing' trend in progressively less massive systems. Ellis et al (1996a) propose a two-component `passively evolving giant plus rapidly evolving dwarf model' as the simplest empirical description of their data.

## *3.3 Other Probes of Luminosity Evolution*

One of the most satisfactory developments in galaxy evolution has been the verification of the redshift survey trends from completely independent approaches and, in particular, from the study of QSO absorbers. A redshift survey of 55 galaxies with 0.3<z<1 producing Mg II absorption in the spectra of background QSOs indicates little change in the volume density, characteristic luminosity or rest-frame colors of typical $L^*$ galaxies with established metallic halos (Steidel et al 1994). Of importance, there appears a marked distinction between these systems and the rapidly evolving later-type galaxies that do not seem capable of producing significant Mg II absorption (Steidel et al 1993, 1996a). Although the straightforward physical interpretation has been questioned by Charlton & Churchill (1996) because of a poor correlation between the galaxy-QSO impact parameter and the Mg II absorption structure (Churchill et al 1996), and the paucity of non-absorbing galaxies found at small impact parameters, these problems may relate to the difficulty of extrapolating from kinematics on small scales to the large scale properties of the absorbing galaxies.

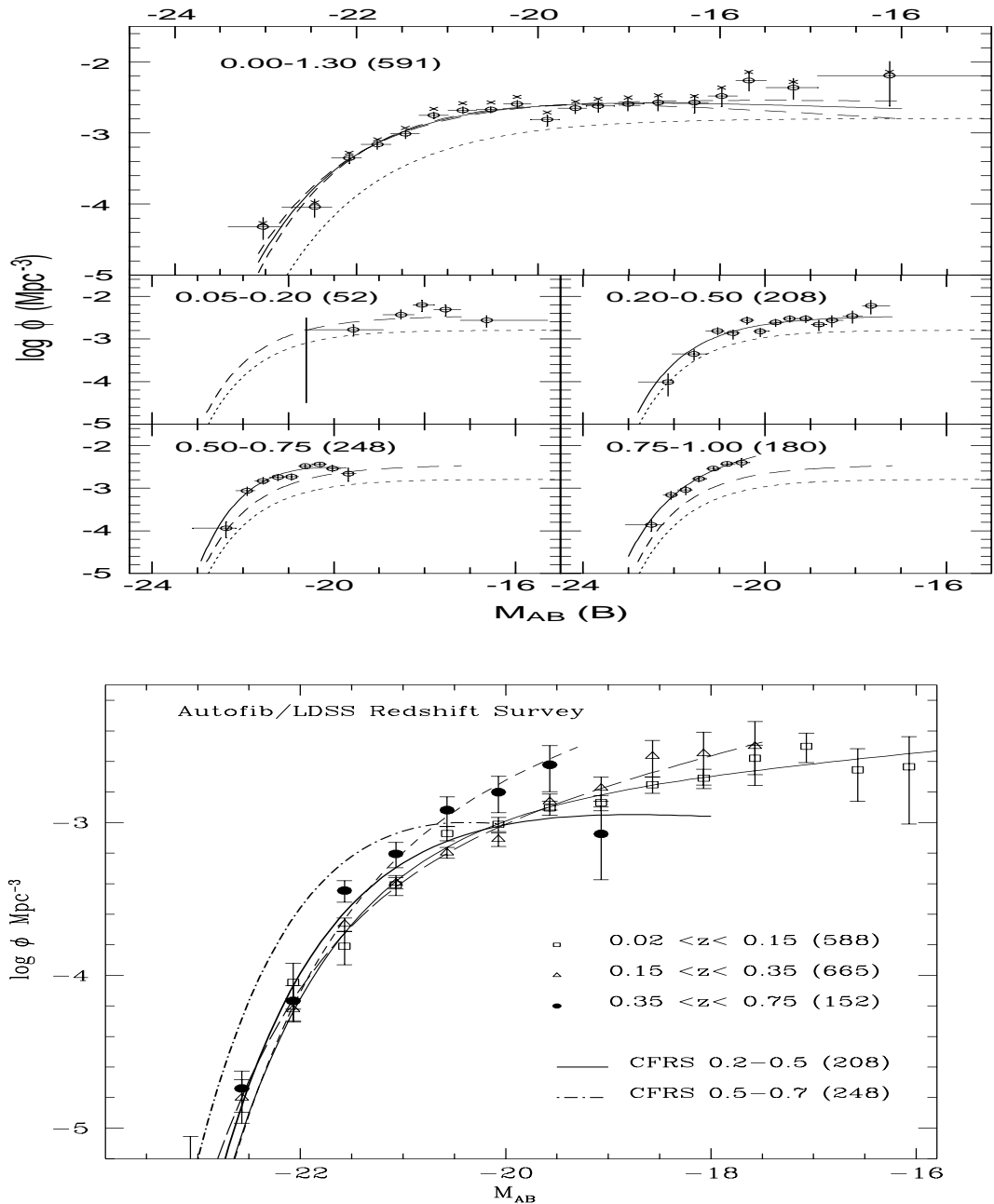

*Figure 6: (a) The rest-frame B(AB) luminosity function for various redshift ranges derived from the CFRS (Lilly et al 1995); numbers in parentheses denote sample size. Solid and dashed curves represent fitted Schechter functions (best-fit and 1-σ fits); the dotted curve represents Loveday et al's (1992) local LF. In each panel where z>0.2, the dashed curve represents the fit for 0.2<z<0.5. With a single photometric scale, clear evidence of luminosity evolution is apparent. (b) Comparison of the Autofib/LDSS (Ellis et al 1996a) and CFRS LFs in the range 0.2<z<0.7 in the CFRS B(AB) system. For convenience both panels adopt $H_o$ =50 km $s^{-1}$ $Mpc^{-1}$ and $\Omega$ =1 as in Lilly et al (1995). Numbers denote sample sizes in the appropriate redshift ranges for the two surveys. The CFRS results admit modest evolution for luminous galaxies whereas the Autofib/LDSS results suggest significant changes occur only for $L<L^*$ galaxies.*

Progress has also been made in studying possible evolution in the surface photometry and dynamics of various classes of distant galaxies. The crucial point in these studies is that evolution is normally inferred via a comparison with local data. Consequently it is important to understand exactly how the distant samples are selected. On the basis of luminosities and scale lengths for a magnitude-limited sample of galaxies selected to have faint bulge components drawn largely from the CFRS, Schade et al (1996) claimed a mean rest-frame B disk brightening of $\approx 1$ mag compared to Freeman's (1970) law by z=0.55. Although this seems difficult to reconcile with LF studies from the same survey (Figure 6a,b), it should be noted the LF trends are $\Omega$-dependent whereas the surface brightness test is not. Moreover, Schade et al applied their test globally to all systems with weak bulges and presumably this incorporated the rapidly evolving lower luminosity systems discussed in Section 3.2.

Vogt et al (1996) obtained spatially-resolved rotation curves for nine galaxies imaged with HST and, using the Tully-Fisher relation, obtained more modest changes of $\approx 0.6$ mag to $z \approx 1$ but warn of possible biases toward more luminous, larger, star-forming galaxies at high z. Rix et al (1997) and Simard & Pritchet (1997) sought to address the question of whether a faint blue galaxy is rotationally supported and thus selected the bluest or strongest emission line sources at z=0.25-0.45 considered to be typical of the excess population. Rix et al measured line widths indicative of rotational support and, as with Simard & Pritchet, claimed to demonstrate that their galaxies are at least 1.5 mag brighter than expected from the local Tully-Fisher relation. The difficulty is finding an appropriate local calibration for these galaxies. Rix et al presented various alternatives but more work is needed to define a self-consistent dynamic dataset over a range in redshift in conjunction with HST images.

## *3.4 Imaging Surveys with HST*

Given the heroic efforts to resolve the faint blue population from ground-based telescopes (Giraud 1992, Colless et al 1994), one might imagine rapid progress in understanding their nature would be possible from the first refurbished HST images that became available in 1994. In practice, interpretation of the HST data has been hindered by a number of factors introduced in Section 2. The following discussion concentrates on post-refurbishment data. Discussion of the HDF is postponed until Section 6.

Two survey techniques have been used to image the faint population. The MDS (Griffiths et al 1994) has utilised parallel WFPC-2 data that sample random high-latitude fields in F785W or F814W and, where possible, F555W. The typical primary exposure time has seriously limited long parallel exposures and thus the bulk of the useful data comes from exposures of $\approx 1$h in F814W. The primary *advantage* of the MDS is the total survey field area. Morphological data and image parameters have been presented to $I_{814}$=22-23 (Glazebrook et al 1995b, Driver et al 1995a) for over 300 sources to $I_{814}$=22 within 13 WFPC-2 fields (0.02 deg$^2$). Driver et al (1995b) also analyse 227 galaxies from a single deeper pointed exposure of 5.7h to $I_{814}$=24.5. The main *disadvantage* of the MDS is that the small field of view of each WFPC-2 image is poorly matched to ground-based multi-object spectrographs and thus follow-up redshift work is rather inefficient. The MDS survey is thus best viewed as a deep 2-D survey of the faint sky. Windhorst et al (1996) provides a good summary of the overall results. The alternative technique discussed by Ellis (1995), Schade et al (1995), Cowie et al (1995a,b), Koo et al (1996) and LeFevre et al (1996a) involves taking HST images in primary mode of ground-based redshift survey fields. By arranging WFPC-2 exposures in a contiguous strip, an effective match is obtained with existing redshift data. A variant here is the exploitation of the `Groth strip', a GTO exposure in F606W and F814W that consists of 28 overlapping WFPC-2 fields (Koo et al 1996).

Glazebrook et al (1995b) and Driver et al (1995a,b) have classified the MDS and related samples visually into spheroidal/compact, spiral and *irregular/peculiar/merger* categories. They claimed the number of regular galaxies (spheroidals and spirals) to I=22-24.5 is approximately as expected on the basis of the local passively-evolving populations, whereas the irregular/peculiar/merger population is considerably in excess of expectations with a much steeper count slope. Abraham et al (1996b) derived the same conclusion from an automated treatment of image morphology based on the concentration of the galaxy light (which correlates closely with the bulge/disk ratio) and the asymmetry (used to locate irregulars). Odewahn et al (1996) investigate the use of artifical neural networks to classify the data and reach similar conclusions. Note that the latter techniques do not yet, as presented, include corrections for bandpass shifting biases. In each of these studies, galaxies were located after smoothing the MDS images to ground-based resolution to avoid double-counting close pairs of galaxies and the conclusions were based on the increased normalisation of the LF discussed earlier.

It is tempting to connect the rapidly-evolving blue galaxies in the redshift surveys with the irregular/peculiar/merger systems seen in the MDS data (Ellis 1995) but central to the MDS results is the physical interpretation of the *irregular/ peculiar/merger* class. Could this category of objects not simply be an increasing proportion of sources rendered unfamiliar by redshift or other effects? A convincing demonstration that this is not the case requires a detailed analysis of the inferred morphology, when viewed at likely redshifts, of a representative multi-color CCD sample of nearby galaxies of known type. Although the machinery is available to conduct such simulations (Abraham et al 1996b; Figure 4), it is not yet clear whether the available local samples are  properly represented in all classes. However,  thus far it seems unlikely such a large bias could occur within the redshift range appropriate for I<22 samples as defined from the CFRS, because the F814W images correspond typically to rest-frame B. However, the precise distinction between late-type spiral and irregular/peculiar/merger may remain uncertain (Ellis 1996a).

Within the limited sample sizes available through HST images of the redshift survey galaxies, the identification of such a high fraction of irregular galaxies is less clear. Schade et al (1995) studied 32 CFRS galaxies with z>0.5 using WFPC-2 images in F450W and F814W and commented on the high proportion (30%) of *blue nucleated galaxies*. Many appear asymmetric and some shows signs of interaction. Cowie et al (1995a,b) discussed the unfamiliar nature of their distant sample and, for the faintest bluest sources with z>1, introduced the terminology of *chain galaxies* - multiple systems apparently merging along one dimension (but see Dalcanton & Schectman 1996). The varied interpretation of these HST images underlines the difficulty of connecting high redshift morphology with local ground-based data. Possibly the most robust conclusion so far is that the bulk of the faint blue population are comparable in size to the remainder of the population; a result apparent in earlier ground-based efforts (Colless et al 1994). Few are truly compact systems of the kind selected for detailed study by Koo et al (1995) and Guzman et al (1996).

### *3.5 Evolutionary Constraints from Galaxy Correlation Functions*

The analysis of positional data in deep galaxy catalogues in terms of 2-D angular correlations or 3-D redshift space correlations is primarily useful as a probe of the evolution of clustering with look-back time (Phillipps et al 1978). The role of correlation functions in constraining *galaxy evolution,* as stressed by Koo & Szalay (1984), has, to a large extent, been overtaken by the redshift surveys, which give a much clearer and less ambiguous indicator of luminosity changes. Nonetheless, a bewildering amount of 2-D data has been analysed in recent years, mostly from panoramic photographic and CCD-based surveys in various bands and the results arising from this work have featured prominently in the debate on faint blue galaxies.

The angular correlation function, $w(\theta)$, of faint galaxies is linked to its spatial equivalent, $\xi(r)$, by terms that depend on the redshift distribution (see Peebles 1994 for definitions). The connecting relationship (Limber 1953, Phillipps et al 1978) takes into account both the angular diameter distance, $d_A = d_L/(1+z)^2$ and dilution from uncorrelated pairs distributed along the line of sight. In practice, as $\xi(r)$ is locally type-dependent (Davis and Geller 1976), $w(\theta)$ is weighted by the type-dependent redshift distribution $N(z,j)$ determined by the k-correction and evolutionary behaviour of each type. A further factor is the likely evolution in spatial clustering conventionally parameterized in proper space as:

$$\xi(r, z) = (r/r_o)^{-\gamma} (1+z)^{-(3+\varepsilon)} \qquad [6]$$

where $r_o = 5\ h^{-1}$ Mpc (Peebles 1980) is the current scale-length of galaxy clustering and $\varepsilon = 0$ corresponds to clustering fixed in proper coordinates. For a correlation function $\xi(r) = (r/r_o)^{-1.8}$, clustering fixed in comoving coordinates yields $\varepsilon = -1.2$ whereas linear theory with $\Omega=1$ indicates growth equivalent to $\varepsilon = +0.8$ (Peacock 1997).

Broadly speaking the angular correlation function results for the faint surveys are consistent with a decline in the spatial correlation function with increasing redshift (Roche et al 1993, 1996b; Infante & Pritchet 1995, Brainerd et al 1995, Hudon & Lilly 1996, Villumsen et al 1996) but any quantitative interpretation that would allow statements to be made on luminosity evolution would have to make careful allowance for the change in apparent mix of types with redshift. Because of the morphology-density relation, the k-correction works in the direction of suppressing the visibility of correlated spheroidal galaxies and therefore an apparent decrease in the amplitude of clustering is inevitable. Using color-selected subsamples, some workers (Efstathiou et al 1991, Neuschaefer et al 1991, Neuschaefer & Windhorst 1995) have addressed this issue and claimed that a large fraction of the faintest blue samples must be weakly correlated unless they are at very high redshift (z>3). The surprisingly large decrease in angular clustering with apparent magnitude has even led to the suggestion that the bulk of the faintest sources represent a population unrelated to normal galaxies (Efstathiou 1995).

The dramatic evolution claimed from the 2-D datasets contrasts somewhat with those studies based on 3-D data from the redshift surveys. Cole et al (1994b) derive spatial correlation functions from the Autofib redshift survey and Bernstein et al (1994) analyse panoramic 2-D data within magnitude limits where the redshift range is likewise known. Neither found significant changes in the clustering scale to z=0.3 other than can be accounted for if the bluer star-forming galaxies were somewhat less clustered as seen locally in IRAS-selected samples. Estimates of spatial clustering at higher redshift in the CFRS redshift surveys give rather uncertain, though modest, growth estimates ($0 < \varepsilon < 2$) (LeFevre et al 1996b). The results from the deeper Keck LRIS survey (Carlberg et al 1997) also indicate modest evolution (a 60% decrease in $r_0$ from z=0.6 to 1.1) with a possible strong cross-correlation between high- and low-luminosity galaxies on 100 $h^{-1}$ kpc scales.

The apparent conflict between the weak angular clustering of faint sources and the modest decline in the spatial clustering to z=1 might be understood if an increasing proportion of the *detected* population is (a) less clustered and (b) itself evolving at a physically reasonable rate. Villumsen et al (1997) argue that over 20<R<29, $w(\theta)$ can be explained via the linear growth of a component whose present correlation length is comparable to that of local IRAS galaxies ($r_o \approx 4\ h^{-1}$ Mpc). Smaller correlation lengths ($\approx 2\ h^{-1}$ Mpc) are claimed by Brainerd et al (1995) but the differences may relate to the adopted redshift distributions for the faint populations as well as possible sampling variations in the small fields available to date. Peacock (1997) concludes the evolution in $\xi(r)$ seen by LeFevre et al can be simply interpreted via a single population of galaxies evolving with $\varepsilon \approx 1$ - corresponding to $\Omega=0.3$. That such diverse conclusions can emerge from these basic observational trends gives some indication of the degeneracies involved in analyzing the data.

R S ELLIS

## 4. ADDRESSING THE MAIN ISSUES

The aim in this section is to distill the observational phenomena discussed in Section 3 into a few key physical issues that will assist in understanding the evolutionary role of faint blue galaxies. As the greatest progress has been made in the redshift interval to z=1, the discussion of fainter samples and the HDF will be deferred until Section 5.

### *4.1 Uncertainties in the Local Field Galaxy Luminosity Function*

The nature of the local LF was comprehensively reviewed by Binggeli et al (1988). Since then, considerable progress has been achieved through deeper, more representative local optical and infrared surveys such as the the Stromlo-APM (Loveday et al 1992), Las Campanas (Lin et al 1996), CfA (Marzke et al 1994), SSRS2 (da Costa et al 1994) and DARS (Peterson et al 1986, Efstathiou et al 1988, Mobasher et al 1993) redshift surveys. These are utilised both as photometric databases for normalising the galaxy counts at the bright end (Maddox et al 1990) and in deriving LFs as a function of type. Colless (1997) provides a recent summary of local LF determinations (Figure 7) to which is added the recent LF analysis of the ESO Slice Project (ESP, Zucca et al 1997), a survey of 4044 galaxies to $b_J$=19.4.

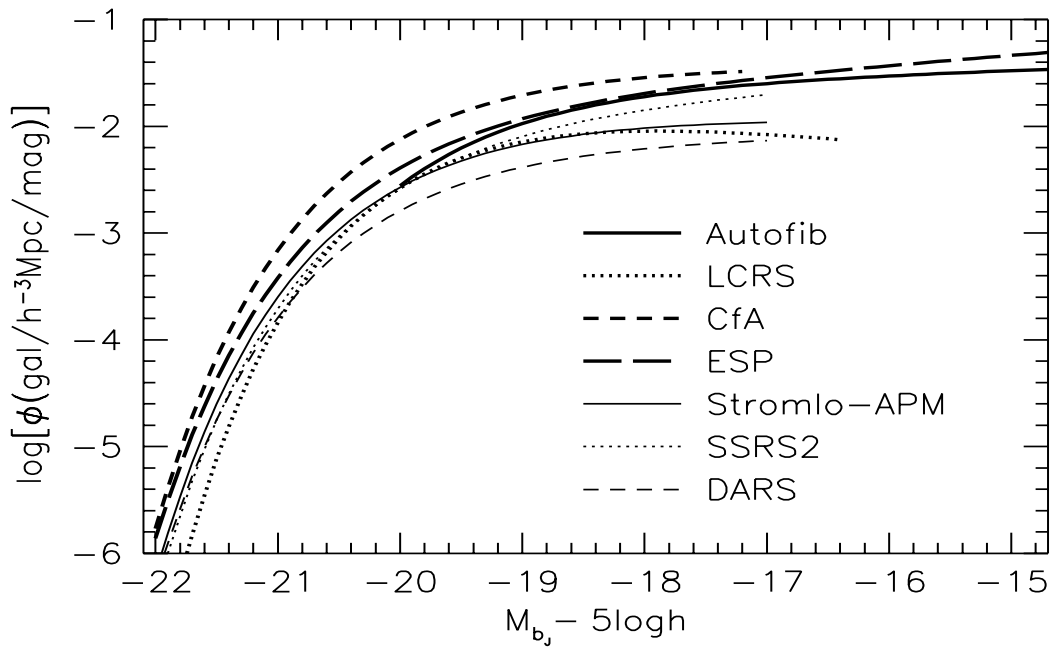

*Figure 7: A comparison of the Schechter function fits to various determinations of the local galaxy LF from the compilation of Colless (1997), updated to include the results of Zucca et al (1997). All LFs are placed on the $b_J$ system using transformations given by Colless (1997).*

The local normalisation depends on a combination of the LF parameters (not just $\phi^*$, see equation [3] ) and its value is central to the question of whether there is significant evolution in galaxy properties to z=0.3-0.5. Early faint galaxy workers (Broadhurst et al 1988, Colless et al 1990) assumed either the Efstathiou et al (1988) or Loveday et al (1992) normalisation, $\phi^* = (1.40\pm0.17)\times 10^{-2}\ h^3\ \mathrm{Mpc}^{-3}$, which, if too low, would increase the evolution inferred assuming $L^*$, $\alpha$ are correctly determined. Although this might be explained via a local minimum in the southern galaxy distribution, there is no convincing indication of this in the redshift distribution and so the anomalously steep count slope found by Maddox et al (1990) brighter than B=18 has been taken to indicate that a problem may exist elsewhere.

The Stromlo-APM survey limited at $b_J$=17.2 is the largest and most well-documented local galaxy survey, but its photometry has been questioned in several recent papers. Metcalfe et al (1995b) and Bertin & Dennefeld (1997) compare APM magnitudes with their own CCD-calibrated measures and detected a scale error and a considerable scatter in the region 17<$b_J$<19. Bertin & Dennefeld explained this, in part, to unmeasured light fainter than the relatively bright APM isophote. However, Metcalfe et al found that high surface brightness galaxies lose more light than those of low surface brightness because of the limited dynamic range of the plate measuring machines involved. Bertin & Dennefeld present B-band counts whose dlogN/dm slope, $\gamma_{16:19}$ in the 16<$b_J$<19 range is 0.53 (cf 0.59 for Maddox et al), a value more consistent with no evolution expectations. It should be noted that any photometric limitations of the APM data may possibly apply also to the COSMOS-based photometry used by Zucca et al (1997) because Metcalfe et al and Rousseau et al (1996) found similar behaviour inCOSMOS data. However, not all photographic photometry may be affected in this way. Weir et al (1995) present photographic counts whose slope is flatter that of Maddox et al.

The above calibrations of non-linear photographic survey data using linear detectors are the forerunners of a new generation of photometric surveys based entirely on CCDs or infrared arrays. Gardner et al (1996) presented counts in BVIK for two large northern fields covering 10 deg$^2$ and found a B band slope $\gamma_{16:19}$ = 0.50±0.03 in reasonable agreement with Bertin & Dennefeld (1997). Huang et al (1997) present counts in BIK for a similar sized equatorial area. Although both Gardner et al and Huang et al claim to disagree, the difference lies mostly in the bright K data. Huang et al's Table 3 gives $\gamma_{16:19}$ = 0.53. Although these survey areas are very small in comparison with the 4300 deg$^2$ of the APM survey, they nonetheless support the above claims for a photometric revision. For the $\alpha$=-1 Schechter function adopted by Loveday et al, Gardner et al's counts indicates $\phi_B^*$=2.02 x 10$^{-2}$ h$^3$ Mpc$^{-3}$ i.e. a normalisation about 50% higher than the APM result. A similar high normalisation emerges from the ESP results limited at $b_J$=19.4 (Zucca et al 1997).

The LF and counts are also available from the extensive Las Campanas Redshift Survey (LCRS, Lin et al 1996). Galaxies were selected in *r* according to a relatively bright isophote over an area of 720 deg$^2$. When allowance is made for the different photometric scale, Lin et al found a low normalisation (as Loveday et al) and somewhat discrepant $\alpha$ and L$^*$ in comparison to those of other surveys. The latter point is significant given that the count normalisation depends on all three parameters. The differences are not understood but may arise from the poor match between their actual LF and the Schechter function, possible inaccuracies in the single type-invariant k-correction used (a criticism also applicable to the ESP survey) and the outer redshift limit at which the final normalisation is made.

In summary, these issues, although not entirely resolved, support a normalisation increase of at least 50% over that adopted by Loveday et al (1992) and indicate that at least *some* photographic data brighter than B=19 may be inaccurate in both photometric scale and possibly even in source detection (see Section 4.2 below). The normalisation derived for galaxies with z<0.1 observed fainter than B=19 (Ellis et al 1996a, Cowie et al 1996) support this upward revision, which has been incorporated in Figure 5(a).

Turning to the *shape* of the LF and its dependence on morphology or spectral class, Driver & Phillipps (1996) have demonstrated the difficulty in estimating the faint end slope reliably from magnitude-limited redshift surveys. Regardless of size, such surveys are optimised to define the LF in the region around M$^*$. An additional difficulty with the deeper LCRS and ESP surveys (Lin et al 1996, Zucca et al 1997) is their reliance on a *single k-correction formula* for all galaxies. If, as suggested by Marzke et al (1994), the mean spectral type is a function of luminosity, systematic errors may be made in the volume correction, leading to a distortion in the derived shape.

Recognising these local uncertainties, Koo et al (1993) and Gronwall & Koo (1995), postulated the existence of a population of intrinsically faint blue star forming galaxies whose LF has a much steeper faint end slope; this serves to minimise the evolution necessary to explain the faint galaxy counts. Limited evidence in support of an underestimated component of blue dwarf galaxies has been presented by Metcalfe et al (1991), Marzke et al (1994) and Zucca et al (1997). The hypothesis was central to the `conservative' model proposed in Koo & Kron's (1992) review.

Although the faint end of the local field LF will inevitably remain uncertain to some unsatisfactory degree, to make a fundamental difference in the interpretation of the faint counts without affecting the redshift distributions at the limits now reached is fairly difficult. At the time of Koo & Kron's review, the faint redshift data only extended to B=22.5. However, by extending the redshift surveys to B=24, Glazebrook et al (1995a), Ellis et al (1996a) and Cowie et al (1996) found only a minimal contribution of local dwarfs to the faint counts. Few 22<B<24 galaxies are observed with z<0.1 in marked contrast to Gronwall & Koo's (1995) predictions (Glazebrook et al 1995a, Cowie et al 1996). Although some incompleteness remains at these limits, it is difficult to understand how this could apply to blue systems at such low redshift where the normal spectroscopic diagnostics are readily visible. Driver & Phillipps (1996) correctly pointed out that such inferences should be drawn from a comparison of *absolute* number of sources seen rather than by matching N(z) distributions normalised to evolving numbers. Even so, taking the z<0.1 redshift survey data fainter than B=17, no convincing evidence exists to support a faint end slope steeper than $\alpha=-1.2$ down to at least $M_B=-15 + 5 \log h$ (Ellis et al 1996a, Cowie et al 1996).

Zucca et al (1997) present evidence for an upturn in their local LF fainter than $M_B=-17 + 5 \log h$, similar to trends observed for cluster LF's (Bernstein et al 1995). Although there is no reason to suppose the field and cluster LFs should be the same, as the faint end of the cluster galaxy LF is more reliably determined (given the volume-limited nature of that data), the suggestion is an important one. However, the upturn identified by Zucca et al is based on 37 galaxies fainter than -16 and only 14 fainter than -15. The implications of such a LF upturn in the apparent magnitude range probed by the redshift surveys would be very small. Thus there is no conflict between this result and the local LF limits provided by the faint redshift surveys. However, a significant contribution to the counts from this upturn would not commence until B=26 (cf model (c) of Driver & Phillipps 1996). The sub-Euclidean number count slope beyond B=25 (Figure 5a) provides a further constraint on such a dwarf population.

In conclusion, much work is still needed to verify the detailed form of the local LF. The large scale redshift surveys underway at the AAT and shortly with the SDSS to B=19-20 will make a big improvement provided their photometric scales are robust. However, despite these uncertainties, the deeper redshift surveys now available indicate that a poorly-understood faint end slope fainter than $M_B=-15 + 5 \log h$ is unlikely to seriously distort our understanding of the faint counts to B≈26.

## *4.2 The Role of Low Surface Brightness Galaxies*

Although intimately connected with uncertainties in the local LF, the presence or otherwise of an abundant population of low surface brightness galaxies (LSBGs) is best considered as a separate issue affecting the interpretation of faint data. McGaugh (1994, 1996) has shown how the presence of systems with central surface brightnesses fainter than $\mu_B=23$ arcsec$^{-2}$ covering a wide range of luminosities would seriously affect determinations of the LF particularly if the local data were plagued by isophotal losses as discussed in Section 4.1. McGaugh (1994), Phillipps &

Driver (1995), Ferguson & McGaugh (1995) and Babul & Ferguson (1996) have explored this uncertainty and proposed the existence of an abundant population of local LSBGs that could be faded remnants of blue star-forming systems identified fainter than B>22. The idea stems from Babul & Rees' (1992) suggestion that the excess seen in the counts might arise from a separate population of dwarfs whose initial star-formation era is delayed until the UV ionising background drops below a critical value. Rapid fading would produce a large present-day abundance of red LSBGs. The role of LSBGs in these suggestions is thus two-fold; first as an additional uncertainty in the local LF and second as remnants of the faint blue galaxies.

Quantifying the contribution of LSBGs to the local LF will remain controversial until a suitable catalogue exists for which rigorous selection criteria have been applied *and redshifts determined*. Most of the available field data is angular-diameter limited from photographic plates (Impey et al 1988, 1996, Schombert et al 1992, Sprayberry et al 1996) and, although illustrating the range of scale-lengths and surface brightnesses possible, cannot easily be converted into volume limited data. McGaugh (1996) convincingly argues that the presence of only a small number of LSB examples implies a significant correction must be made to the faint end slope, although several assumptions are made in calculating the survey volume as a function of limiting surface brightness.

Although CCDs have been used for LSBG searches in clusters, Dalcanton et al (1997) have analysed 17.4 deg$^2$ of deep transit scan CCD data and identified seven LSBGs with $\mu_V > 23$ arcsec$^{-2}$. Of importance, spectroscopic data for this sample have provided an estimate of the volume density of LSBGs of known physical size. The large mean distance for these LSBGs indicates a LSBG volume density, though still uncertain, that is comparable to that of normal galaxies. As expected, the contribution to the overal luminosity density is very small. Dalcanton et al (1997) argue, as McGaugh (1996), that part of the normalisation change in the local LF (discussed in Section 4.1) may arise from the selective loss of these systems in the bright photographic data.

Although LSBGs undoubtedly exist and may be quite abundant with a range of properties, it seems unlikely that they dominate the luminosity density or that the bulk of them represent the faded remnants of a faint blue population. The fading required to push a typical faint blue galaxy below typical local detection thresholds would have to be 2.5-4 mag, depending on the true faint end slope of the local LF (Phillipps & Driver 1995). To be effective, given the narrow time interval involved, the end of the star-forming era would also have to be very abrupt. Moreover, most well-studied LSBGs are gas-rich and blue (de Blok et al 1996) and quite unlike the postulated faded remnants. Neither is there an obvious correlation between LSBG central surface brightness and color (McGaugh & Bothun 1994). An abundant population of faded remnants would be detectable as a significant Euclidean upturn in the number of red and infrared sources at faint limits which has not yet been seen (Babul & Ferguson 1996).

### *4.3 Is Number Evolution Required?*

Are simple number-conserving models adequate? Or are additional populations of star-forming sources required with no detectable local counterpart as has been suggested in hypotheses based on a postulated population of fading dwarfs (Babul & Rees 1992, Cole et al 1992) or recent galaxy-galaxy mergers (Rocca-Volmerange & Guiderdoni 1990)? Although the upward revision in the normalisation of the local LF and the possibility of a steeper faint end slope of blue sources discussed in Section 4.1 reduces the no evolution N(m,z) paradox originally introduced by Broadhurst et al (1988), even internally within the CFRS data (Lilly et al 1995) there are signs of strong evolution from z=0.3 to 1, i.e. neglecting any reference to local data. Similar conclusions can be drawn from the internally-presented LFs in the Ellis et al (1996a) and Cowie et al (1996) samples although, as these data comprise a number of individual surveys with slightly different photometric selection criteria, the conclusions are perhaps less compelling.

To understand the extent to which this evolution is number-conserving, it is important to revisit the nature of the LF changes presented in Figures 6 in the context of simple, number-conserving, luminosity evolution. Whereas Lilly et al (1995) claim at most a 0.5 mag brightening in the luminosity scale of their redder population over 0.3<z<0.6, they detect a brightening of more than one mag in the luminosity scale for the bluer population in the same interval. Further brightening is seen at high redshift, particularly for less luminous sources. Because of the limited magnitude range sampled in the CFRS, the luminosity overlap at various redshifts is relatively small but, given the homogeneity of the survey, the data by Lilly et al provide incontrovertible evidence for evolution in the blue sources corresponding to a factor 2-2.5 increase in the comoving rest-frame 4400 A luminosity density over 0.3<z<0.8 depending on $\Omega$ (Lilly et al 1996).

Although the Autofib/LDSS redshift survey does not penetrate as deep as the CFRS, the large number (548) of fibre-based redshifts in the intermediate interval 19.7<$b_J$<22.0 (selected primarily from 4-m prime focus plates (that are presumably unaffected by the photometric difficulties discussed in Section 4.1) is a valuable component of this survey which enables the LF to be probed at all redshifts in a consistent manner to $M_{bJ}$=-18 + 5 log h. By virtue of the blue selection, the overall LF is more affected by the evolutionary changes than in those selected in I and K. Ellis et al (1996a) reject an unchanging local LF as a fit to their data to z=0.75 with very high significance. Furthermore, statistical tests reveal that the evolution begins strongly beyond z=0.3 in the sense that there is a steeper faint end slope; no convincing shift is seen in $L^*$ for the overall population. As with the CFRS survey, the bulk of these changes occur in the star-forming component selected with strong [O II] emission. The luminosity decline or fading of the [O II]-strong population is roughly comparable to that seen for the blue galaxies in CFRS.

The question of whether the observed changes in the LF can be understood principally as a luminosity shift for some subset of the population (*pure luminosity evolution [PLE] models*) or whether more complex models are required (e.g. ones that violate number conservation) is difficult to determine from LF data alone because of the statistical nature of the observations. A galaxy may enter the [O II]-strong or blue class only temporarily and thus a variety of physical scenarios might be compatible with the trends observed. However, although there seems a natural reluctance amongst some workers to contemplate `exotic' interpretations of the data (Shanks 1990, Koo et al 1993, Metcalfe et al 1996), luminosity-*in*dependent evolution could be even harder to understand physically, particularly when extending such models to very high redshift.

The traditional PLE models date from Baade (1957) and were explored in detail by Tinsley (1972, 1980) and assume, for simplicity, that all galaxies of a certain class change their luminosities with redshift by the same amount in magnitudes irrespective of their luminosity. Although the LF would maintain the same shape at all redshifts, because different sub-classes are allowed to evolve at different rates with different LF shapes, the integrated form may show evolution that is luminosity-dependent (as observed). In such models (e.g. Pozzetti et al 1996), evolutionary corrections are taken from synthesis codes that aim to reproduce the present-day SEDs via star formation histories of a simple form (Bruzual 1983, Bruzual & Charlot 1993). An unavoidable corollary of those models that can successfully account for the present range of Hubble sequence colors is much stronger evolution for early-type galaxies than for the later types. According to Pozzetti et al, a significant proportion of the excess blue B=25 population are distant young ellipticals with redshifts generally higher than the surveys indicate.

The PLE models represent a subset of an entire class of models whose evolutionary codes aim to satisfy the joint number-magnitude-color-redshift datasets. They can be considered as *ab initio* models (Section 2). Many fit the data to some degree but do not define a unique interpretation.

An important conclusion from the LFs discussed in Section 3 is *the qualitatively different pattern of evolution* that is seen compared to that expected in the standard PLE models. Luminous early-type galaxies appear to evolve very little to z=1 and the bulk of the bluing observed appears to occur by virtue of an increased proportion of star-forming $L^*$ galaxies, few of which have spheroidal forms.

Although number evolution cannot be verified to the limits of the redshift surveys, the situation at fainter limits is much clearer. Figure 8 updates a useful diagram originally published by Lilly et al (1991). Adopting a local LF upwardly normalised with a steeper faint end slope $\alpha$=-1.3 to take into account possible uncertainties discussed in Section 4.1, the calculations in the figure show the maximum redshift to which a source that is brighter than a given luminosity would have to be visible so as to account for the integrated number of sources to B=28 as defined using by the ground-based data of Metcalfe et al (1996). Because the majority of the faintest sources cannot lie at very high redshifts by virtue of the absence of a Lyman limit (Section 5), and it is unreasonable to suppose low luminosity galaxies could shine continuously for a Hubble time, strong number evolution seems unavoidable to account for the integrated population beyond the break at B$\approx$25.

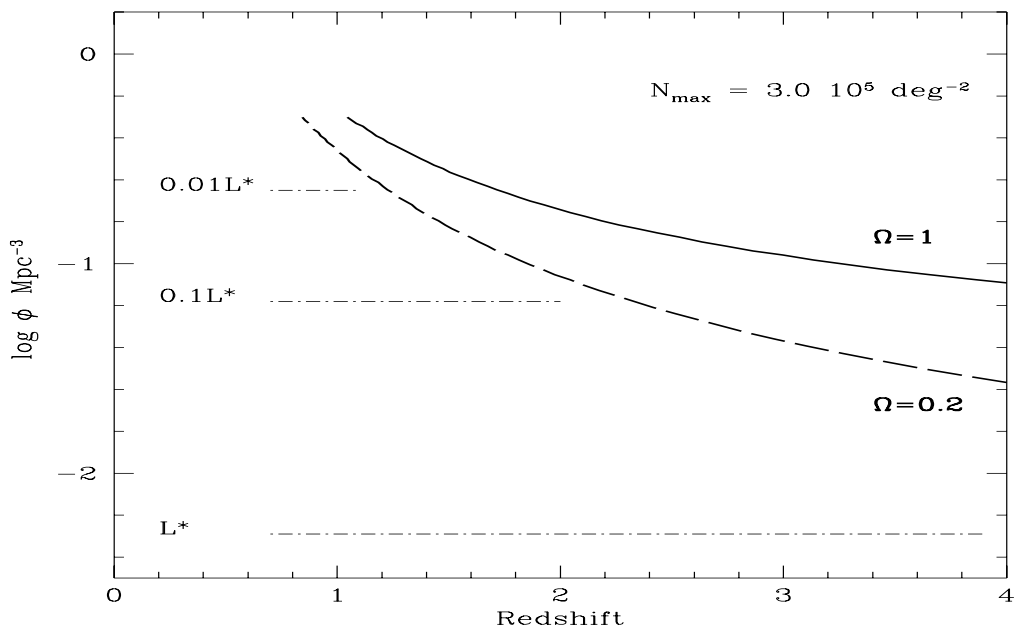

*Figure 8: Evidence for number evolution in the deepest counts. The figure shows the redshift to which a population of galaxies brighter than a given luminosity would have to be seen to account for the integrated number of sources seen in ground-based images to B=28 (Figure 5a). The calculation adopts a high normalisation $\alpha$=-1.3 local LF similar to that observed at z=0.3 in the Autofib/LDSS survey (Ellis et al 1996a).*

## *4.4 A distinct population of rapidly fading objects?*

The difficulties encountered with the traditional PLE models (Section 4.3) can be overcome by invoking a separate population of blue objects that undergo strong luminosity evolution (Phillipps & Driver 1995). Unlike the traditional PLE models, this hypothesis invokes a delayed star formation in a subset of the galaxy population followed by a remarkably rapid decline in activity thereafter. The present-day remnant of this population would presumably occupy the faint end of the local LF. Although physical models have been proposed to support this picture (Cowie et al 1991, Babul & Rees 1992, Babul & Ferguson 1996), the original hypothesis was proposed only to fit the data. Nevertheless, there is a strong theoretical motivation because *ab initio* models tend to overpopulate the faint end of the local LF (Cole et al 1994a). Evidence in support of this empirical picture includes the very different evolutionary behaviour of field LFs characterised by color and spectral features (Lilly et al 1995, Heyl et al 1997) and the rapid increase in the number of irregulars in the MDS data (Glazebrook et al 1995b, Driver et al 1995b). A high fraction of the HST irregulars are also blue [O II] -strong sources (LeFevre et al 1996a). However, the key to testing this model further would be to verify that a distinct population exists and to demonstrate that it fades according to the rate determined from the LF studies.

Some progress might be made by studying the supposedly rapidly evolving component in more detail, as well as by finding local examples. A typical, rapidly-evolving blue galaxy at z>0.3 has $M_B \approx -19.5 + 5 \log h$, $B-I \approx 1-1.5$, a rest-frame equivalent width of [O II] of 20-40 A and an irregular appearance with knots and occasionally a compact core. Although a few giant irregulars are found locally (Gallagher et al 1989), the bulk of the blue sources with comparable [O II] strengths at low redshift are gas-rich metal-poor dwarfs believed to have suffered sporadic bursts of star formation (Telles et al 1997, Gallego et al 1995, Ellis et al 1996a). Leitherer (1996) provides a comprehensive review of the physical processes responsible for sudden star formation in galaxies. There are two broad categories: (i) H II region-like spectra superimposed on that of an older population (Telles et al 1997) and (ii) nuclear starbursts. Both may be induced by interactions.

Surprisingly little is known about the astrophysical properties of the faint blue galaxies, which at first sight seems remarkable given spectra are available for several thousand such galaxies! With the exception of Broadhurst et al's (1988) original data taken at a spectral resolution of 4 A (which they interpreted as supporting signs of short-term star formation in the distant population), both the signal/noise and resolution of more recent, fainter, data are optimised for little more than measuring redshifts; a resolution of 20-40 A is more typical. Heyl et al (1997) and Hammer et al (1997) discuss the spectral properties of co-added spectral datasets in the Autofib/LDSS and CFRS surveys respectively but disagree on the degree to which short-term star formation may be occurring in their samples. Higher quality spectra are required to make progress.

A tentative connection between the faint blue population and active galaxies is reported by many groups. On the basis of emission line ratios, Tresse et al (1996) claimed between 8 and 17% of the z<0.3 CFRS sample could be Seyfert 2 or LINERs, depending on uncertain corrections for underlying stellar absorption. This is about 4 times higher than the local rate and would suggest a dramatic evolution in the proportion of active sources over quite recent epochs. Likewise, Treyer & Lahav (1996) claimed B<23 galaxies with z<0.3 could contribute 22% of the 0.5-2keV X-ray background. At fainter limits, Windhorst et al (1995) find the microJy radio sources overlap significantly with the faint blue population. Windhorst et al argue that the non-thermal activity is more likely a by-product of interactions and mergers rather than the direct output of classical active galactic nuclei.

How might fading in a given population be demonstrated convincingly? We return to the question of selecting some physical parameter that does not change in the process. Koo et al (1995) and Guzman et al (1996) have studied a particular subclass of compact narrow emission line galaxies (CNELG) selected on the basis of very small effective radii (1kpc) and have suggested that these may be the ancestors of local dwarf spheroidals. Although only a minority of the faint blue population lies in this category, this suggestion is important because, if correct, it would demonstrate for a subset of the faint population that significant fading has occurred. Spectroscopic velocity dispersions from the Keck HIRES are typically 30-50 km s$^{-1}$ which indicates low mass/light ratios and bursting star formation histories. Substantial fading (4-7 mags in $M_B$ in 3 Gyr) would be needed to reduce such systems to the present luminosities of dwarf spheroidals.

The CNELG study illustrates an important way forward because, if local dwarf galaxies are faded versions of their high z cousins, the surface brightnesses within some fixed metric scale must also have declined by the same amount as indicated in the LF studies. The crucial test for the fading hypothesis are thus to examine the surface brightness distributions of various subsets of the field population taking due care to allow for possible selection biases. The availability of HST images for subsets of the large redshift surveys (Ellis 1995, Schade et al 1995, LeFevre et al 1996a) makes this an important direction for future study.

## *4.5 Evidence for mergers?*

An increase with redshift in the rate of galaxy merging is an attractive way to satisfy the changing shape of the LF. Such an explanation is a natural consequence of hierarchical pictures (Carlberg 1992) and features prominently in many of the *ab initio* models discussed earlier (Baugh et al 1996).The importance of merging in the faint counts has a long but somewhat inconclusive history (see Carlberg 1996 for a recent summary). The optical LF data has been particularly difficult to interpret in this picture, primarily because the rest-frame light indicates the history of *star formation* whereas the merger predictions are based on the evolution of the *mass*. Rocca-Volmergange & Guiderdoni (1990) introduced a self-similar, mass-conserving evolutionary model in which the comoving number density is required to increase as $(1+z)^{1.5}$ in a $\Omega=1$ Universe. Broadhurst et al (1990), analysing redshift and number count data in the context of a low local normalisation, required a mass growth rate such that a typical galaxy became 4-6 fragments by z=1. Eales (1993) incorporated a more physically-based model and predicted changes in LF shape not dissimilar to those seen in the recent redshift surveys. To overcome uncertainties in relating mass and light, Broadhurst et al (1990) advocated conducting deep K-limited surveys to test the merger hypothesis and predicted a turnover in the mean redshift at faint limits consistent with the absence of large mass objects at high z. Detailed predictions in this context have been made by Carlberg & Charlot (1992) and the latest available K-selected data provide some support for such a picture (Cowie et al 1996); over K=18-20 the mean population redshift hardly increases with apparent magnitude. However, incompleteness in the z>1.5 range remains a concern.

A more direct approach might be to estimate the interaction rate by searching for close pairs. Barnes & Hernquist (1995) discussed the importance of `major' and `minor' mergers with respect to the morphology of the host galaxy and energy arguments suggest a strong redshift dependence in the interaction rate (Carlberg 1996). However, even the local interaction rate remains uncertain because not all of the diagnostic features of a merger are expected to be easily visible (Mihos 1995). One of the earliest observational studies attempting to define the merger rate at large lookback times was that of Zepf & Koo (1989) who found 20 close pairs in a faint photographic sample limited at B=22 and concluded that the pair merger rate increases with redshift as $(1+z)^m$ where m=2-4. The difficulty here lies in correcting for a large number of biases that may artificially raise the apparent interaction rate. Such biases include the dissimilar tactics in analysing low and high redshift data, a possible boosting in luminosity of satellite galaxies whose star formation is triggered by a merger, and the k-correction, which leads to an increase in the number of late-type spirals at high redshift that often have peripheral H II regions; the latter are rendered more visible in the rest-frame UV. A recent analysis using ground-based data is discussed by Woods et al (1995).

The arrival of HST images has led to renewed interest in this area. Neuschaefer et al (1997) present a comprehensive analysis of the number of close pairs (<3 arcsec) to I=23.5 in 56 MDS fields and analyse their results in the context of HST and ground-based angular correlation functions. In hierarchical merging, one need not expect to find a significant excess in the angular correlation function on small scales. Neuschaefer et al's data supercedes the earlier MDS study of Burkey et al (1994) (which claimed an increase in the merger rate comparable to Zepf & Koo 1989) and gave m = 1.2 ± 0.4, as did Woods et al (1995). No convincing excess of pairs with separations less than 3 arcsec was found in comparison to an extrapolation of the angular correlation function.

Greater progress will be possible when redshifts are available. The physical scale around each host galaxy can then be defined and, more importantly, satellites can be constrained to include only those above a fixed luminosity limit (providing some estimate of the k-correction is made). Patton et al (1997) use the field galaxies located via the CNOC1 cluster galaxy survey (Carlberg et al 1996) to define a sample of close pairs within 20 $h^{-1}$ kpc. Redshifts are available for half of the secondary images and demonstrate that a sizeable fraction are true physical associations. By z=0.33, 4.7±0.9% of the faint population are claimed to be merging and comparison with local data suggests m=2.8±0.9. A number of corrections are required in this analysis to allow for the idiosyncrasies of the observing strategy that produced the primary sample and contamination from cluster galaxies in the secondary sample. LeFevre et al (1996a) have both the advantage of HST images and a very wide redshift range which means that the evolutionary trend can be examined without reliance to any local data. Moreover LeFevre et al sampled to much fainter limits around each primary galaxy. More than half of the major mergers in the LeFevre et al sample have z>0.8; indeed no strong trend is seen until the redshift is quite large, in qualitative agreement with the predictions of hierarchical clustering (Baugh et al 1996). However, the absolute rate remains uncertain when determined by pair counts and so whether merging is the dominant process driving the evolution of the LF remains unclear.

The above studies demonstrate the difficulties of verifying the merger hypothesis quantitatively. However, some important points can be made. First, following the upward revision in the local LF (Section 4.1), there is less need for rapid number evolution at low z and this considerably reduces the difficulties concerning the abundance of recent merger products (Dalcanton 1993). Second, notwithstanding the uncertainties, there is growing observational evidence from HST images of galaxies of known redshift and the modest depth of the faintest K-limited surveys that merging is of increasing importance at high redshift.

## 5. SURVEYS BEYOND THE SPECTROSCOPIC LIMIT

Various workers have considered ways in which the mean redshift of the very faint population (say B≈27, R≈26, I≈26, i.e. beyond normal spectroscopic limits) could be estimated statistically. One expectation might be that the bulk of such a faint population lies beyond z=1. A redshift of unity has represented a significant barrier to the systematic study of normal galaxy populations for many years. It corresponds to the 4-m telescope limit for which spectroscopic redshifts are possible for normal $L^*$ field galaxies and it also marks an important transition from analyses that are, broadly speaking, independent of the cosmological framework, to ones where the volume element and time-redshift relation depend critically on $\Omega$ and $\Lambda$. A lesson that emerges from Section 4 is that a fundamental difficulty in making progress is the need to ensure that local and high z datasets are treated similarly. As we turn to the next logical step in the observational challenge, this becomes even more the case.

### *5.1 Constraints from Gravitational Lensing*

The suggestion that the mean redshift of the B=27 and I=26 population might be rather low, consistent with discussion of the break in the count slope in Section 3, first arose from the pioneering CCD exposures conducted by Tyson and collaborators (Tyson 1988, Guthakurta et al 1990). Tyson also developed the first practical applications of gravitational lensing by rich clusters as a tool for estimating the mean statistical distance to the background population (Tyson et al 1990).

At faint magnitudes, lensing by foreground masses affects source properties by an amount that depends on the nature of the intervening lens, the relative distances to the lens and source, and the cosmological model (Blandford & Narayan 1992, Fort & Mellier 1994). The phenomenon manifests itself in several ways depending on the geometrical configuration and lens scale. In the case of *giant arcs*, dense concentrated clusters of galaxies beyond z=0.1 magnify faint sources considerably, extending spectroscopic and photometric detections to fainter limits. Certain clusters have well-constrained mass distributions, either from giant arcs and multiple images of known redshift, or from indirect probes such as X-ray luminosities and velocity dispersions. The lensing shear field viewed through these clusters can provide a statistical estimate of the mean distance to sources that are too faint for conventional spectroscopy. As the technique is purely geometric in nature, it provides an independent probe of the distances to faint galaxies.

Recognising the need to separate the dependence of the weak lensing signal on both the relative distances of source and lens and the nature of the lensing cluster, Smail et al (1994, 1995) compared the shear seen in a background population to I=25 as measured through three clusters at different redshifts. The relatively weak shear found by Smail et al through an X-ray luminous cluster at z=0.54 suggested a relatively low mean redshift for the population but this is in marked contrast to the conclusions of Luppino & Kaiser (1997) who detected a significant shear through a similar cluster at z=0.83. Difficulties arise because this technique relies critically on measuring the *absolute* shear as well as understanding the properties of the lensing cluster. A comparison of the various techniques used to estimate the shear by different workers on the same clusters is badly needed and deep HST data of more distant clusters is also required to correct for seeing and other effects that may affect ground-based images.

Considerable progress is possible if redshifts are available for some of the lensed features, because this reduces the dependence on absolute measurements of the lensing signal (Fort & Mellier 1994). Giant arcs are strong lensing events of high magnification but offer a rather unreliable glimpse of the high redshift population. Spectroscopy is now available for about 20 cases; a few have redshifts beyond 1. Although these are galaxies found serendipitously by virtue of their location behind unrelated foreground clusters and their unlensed magnitudes are quite faint, important selection effects operate in their recognition and thus they are an unreliable statistical probe of the background redshift distribution. Strong lensing is optimal when the background source is around 2-3 times the angular diameter distance of the lens and so a low frequency of high redshift arcs may simply reflect the paucity of concentrated high z clusters. At the moment, few convincing arcs have been seen in clusters beyond z=0.5. Secondly, the arcs are, without exception, found as high surface brightness features in optical CCD images often by virtue of their contrasting blue color as compared to the red cluster population. Smail et al (1993, 1996) have examined the optical-infrared colors and HST angular sizes of giant arc samples and, not surprisingly, deduced that many are representative of late type galaxies undergoing vigorous but extended star formation.

The more important role of arc spectroscopy is to further constrain the properties of the lensing cluster. The lensing geometry can then be used together with the image shapes and orientations for the fainter population of less distorted `arclets' to yield a statistical redshift for each one assuming, on average, that they are intrinsically round sources. This *lensing inversion* technique was first developed by the Toulouse group (Kneib et al 1994) using ground-based data for the well-studied cluster Abell 370, and it has now been extended using HST images and more comprehensive ground-based spectroscopy for Abell 2218 and Abell 2390 (Kneib et al 1996). These studies suggest that the mean redshift of I<25.5 populations cannot significantly exceed 1. In several cases, the constraints on the arclet redshifts are sufficiently tight that the predicted redshifts are worth verifying spectroscopically. In a number of such cases, Ebbels (1996), Ebbels et al (1996) and Bezecourt & Soucail (1997) present convincing evidence that the *inversion* technique works well. However, the construction of genuine magnitude-limited samples appears to be difficult using this method. Not only are the lensed images magnified by different amounts but inversion is unreliable for the smallest sources that increasingly dominate the faint counts (Roche et al 1996).

To overcome difficulties inherent in analyses of distorted images, Broadhurst (1997) proposes the use of *magnification bias* or gravitational convergence. A lensing cluster enlarges the background sky and this produces a diminution in the surface density of sources depending on the relative distances involved. For galaxy counts, the effect is in the opposite direction to the magnification described above (Tyson et al 1984, Broadhurst et al 1995). The background source counts viewed at radius r from the centre of a foreground lens become:

$$N(<m, r) = N_0 (<m) \mu (r)^{2.5 \gamma - 1} \qquad [7]$$

where $N_0 (<m)$ represents the true counts, $\mu (r)$ is the magnification at angular radius r from the center of the lens, and $\gamma=dlogN/dm$ is the slope of the number-magnitude counts. For $\gamma=0.4$, the magnification and dilution effects cancel out and no effect is seen. However, when $\gamma< 0.4$, the counts decrease, particularly near the critical angular radius. The location of this point depends on the relative angular diameter distances of the sources and the lens, as well as on the cosmological model. Fort et al (1997) and Mellier (1996) describe promising applications of this technique. As the method relies only on source counting rather than a reliable measurement of image shapes, the technique can probe to very faint limits in a controlled manner.

## *5.2 The HDF*

The HDF (Williams et al 1996) and related Keck spectroscopic programmes (Steidel et al 1996c; Illingworth et al 1996, Cohen et al 1996, Lowenthal et al 1997) have already given a tremendous boost to studies of the Universe beyond the limits of the 4-m telescope redshift surveys. The investment of a large amount of HST Director's time enabled the first comprehensive *multi-color* study of a deep field with WFPC-2. The relatively poor efficiency of HST blueward of 500 nm had previously prevented even the most adventurous workers from attempting deep UV and blue imaging in normal guest observer allocations. Furthermore, the public availability of the HDF image has led to a rapid delivery of scientific results concerning the nature and redshift distribution of galaxies well beyond spectroscopic limits.

The most striking result, from a consideration of the high surface density of B=28 galaxies in ground-based data (cf Metcalfe et al 1995b, 1996), (Figure 5a), is the large fraction of blank sky in the HDF image. This simple observation reflects the fact that the bulk of the I>25 population has very small angular sizes continuing trends identified first in the MDS (Mutz et al 1994, Roche et al 1996a). Another result that represents a continuation of MDS work is the increasing proportion of faint irregular structures from I=22 to I=25 (Abraham et al 1996a). Although the widely-distributed HDF color images emphasise irregularity via blue features which sample the restframe UV, the morphological analysis of Abraham et al is based on the F814W image which, at $z \approx 1$-1.5, is equivalent to rest-frame U or B. Beyond I=25, the resolution is insufficient for detailed study; indeed, Colley et al (1996) raise the important question of the definition of a galaxy in this regime. The small angular sizes and high abundance compared to z<1 LF estimators, together with the declining slope of the counts (Figure 5a) suggest that many of these sources may be sub-galactic components at an early stage of formation.

The availability of images in four passbands, supplemented by deep ground-based data at infrared wavelengths (Cowie 1996), has led to a surge of interest in estimating redshifts from colors. In its most elementary form, a set of discontinuities (Lyman or Ca II 4000 A break) is located in individual galaxies via imaging through a set of filters. The HDF observing strategy was chosen to extend earlier work by Guhathakurta et al (1990) and Steidel & Hamilton (1992) and isolate those sources whose Lyman limit discontinuities are redshifted into the optical. Similar techniques have also been used via ground based images to detect higher redshift sources around QSOs (Giallongo et al 1996). The validity of the technique is reviewed comprehensively by Madau et al (1996) in the context of what is known about the UV SEDs of galaxies and the attenuation of UV light by intervening HI clouds. They claim the strategy is robust to quite considerable uncertainties in the precise shape of the SEDs near the Lyman limit and possible effects of dust.

Lists of candidate high z galaxies in the HDF selected on the basis of the Lyman discontinuity were first published by Abraham et al (1996a) and Clements & Couch (1996) but most of the progress has been achieved through spectroscopic exploitation of the technique in the HDF and other fields using the Keck telescope (Steidel et al 1996a,b, 1997; Giavilisco et al 1996, Lowenthal et al 1997). The spectra demonstrate a success rate of 100% for the selection of high z sources ( not a single redshift has been confirmed outside the expected range). The Keck surveys are still in progress but already provide a significant constraint on the proportion of R<25 star-forming galaxies beyond z=2.3 (the redshift corresponding to the limit entering the F300W filter). These results have offered a valuable glimpse at the nature of a population of galaxies with 2.3<z<3.5 and perhaps the most significant results are their relatively modest star formation rates (SFR, 1-6 $M_o$ yr$^{-1}$ ) and volume densities comparable to those of local L$^*$ galaxies ($\approx$ 8 10$^{-4}$ h$^3$ Mpc$^{-3}$) (Steidel et al 1996a,b, Madau et al 1996)[1] . Comparable star formation rates for high z galaxies have been estimated from independent studies by Ebbels et al (1996), Djorgovski et al (1996) and Hu & McMahon (1996). The weak emission lines in these high z star-forming galaxies may explain the null results of many years of primeval galaxy searching based on the assumption of intense photoionized Lyman $\alpha$ emission (Djorgovski & Thompson 1992).

A minimum signal to noise ratio in the UV-optical SED is required to convincingly detect the presence or otherwise of the Lyman limit and thus the question remains as to whether a larger fraction of the sources fainter than R=25 have higher redshift. As the longest wavelength band (F814W) still samples the rest-frame UV at z>4, only star-forming objects above some threshold can be visible at high z in the HDF. To eliminate a large population of sources with low SFRs would require K-band imaging to much deeper limits than is currently possible (Cowie 1996, Moustakas et al 1997).

The multi-color HDF data has also been analysed by numerous workers with respect to various template SEDs in an attempt to secure statistical redshift distributions (Table 2). The critical uncertainty in these studies is the form of the UV SED sampled by the optical HDF data for z>2. Lanzetta et al (1996) grafted large aperture optical SEDs from Coleman et al (1980) with much smaller aperture UV data from Kinney et al (1996), whereas the other workers generally adopted model SEDs in the UV. Gwyn & Hartwick (1996) used model SEDs throughout. With the exception of Gwyn & Hartwick, all fitting thus far was done on the basis of *present-day* SEDs in the (unjustified) hope that an evolving SED must somehow move along the locus of those observed today. Although Bershady (1995) discussed this assumption in the context of six-color low z data, it remains unclear what systematic effects this will have at high z with four-color data sampling the UV.

A further difference among the observers is the aperture they have used to measure the colors. Sawicki et al (1997) determined an `optimum aperture' that varies for each source; Gwyn & Hartwick (1996) used a fixed aperture of 0.2 arcsec, whereas Mobasher et al (1996) experimented with apertures of 0.5 - 3 arcsec depending on magnitude. Given the signal/noise and irregular structure of the images, it seems reasonable to expect that the results obtained may depend on the chosen aperture (FDA Hartwick, private communication).

---

[1] *However, note that these figures refer to a $\Omega=1$, $\Lambda=0$ world model and would be much larger if $\Omega<<1$.*

**Table 2: Color-based Redshift Surveys in the HDF**

| Reference | Magnitude Limit | Template SED |
|---|---|---|
| Gwyn & Hartwick (1996) | $I_{814} < 28$ | Bruzual-Charlot models |
| Lanzetta et al (1996) | $I_{814} < 28$ | z=0 observed + Ly$\infty$ |
| Mobasher et al (1996) | $I_{814} < 28$ | z=0 observed |
| Sawicki et al (1997) | $I_{814} < 27$ | z=0 observed + UV models + Ly$\infty$ |
| Cowie (1996) | H+K<22.5 | z=0 observed |

As an illustration of the uncertainties arising from analyses that differ only in the aperture, fitting algorithm and template SEDs used, the redshift estimates of Mobasher et al (1996), Lanzetta et al (1996) and Cowie (1996) are compared in Figure 9. Cowie (1996) tabulates redshifts for a significant sample of K<20 galaxies from the compilations of groups at Caltech and the University of Hawaii. When compared with the photometric redshifts, the agreement is only satisfactory for the optical+infrared data analyzed by Cowie; a surprising fraction of spectroscopically confirmed low z galaxies are considered to be at high z by Lanzetta et al (Figure 9a). Nonetheless, for the H+K<22 sample discussed by Cowie (1996), the *distribution* of photometric redshifts obtained by Mobasher et al and Lanzetta et al are quite similar (Figure 9b). Cowie's distribution reveals a somewhat unphysical gap for 1<z<2 with what seems like an artificial peak coincident with the location of the Lyman limit at the F300W filter.

I=26 is a convenient deeper limit as it represents that faintest I magnitude at which the signal-to-noise ratios for virtually all sources exceed $3\sigma$ in F300W. Figure 9(c) compares the redshift estimates of Mobasher et al (1996) and Lanzetta et al (1996) to this limit. Systematic differences are clearly present with effects similar to those seen in Figure 9(a). Nonetheless, again the overall distributions agree remarkably well (Figure 9d). Of course, agreement between the various workers does not necessarily imply the results are correct. Systematic errors could be introduced by the effects of dust, inaccurate UV SED slopes or very strong emission lines. In this context, that the mean slope of the rest-frame UV SED for the Steidel et al (1996b) galaxies with z>2.3 is less steep than the models predict is worrisome. Dust appears to be an unlikely explanation given the weak HI content. However, in delineating the overall fraction of faint galaxies with z<1, 2 and 3, the photometric redshift techniques appear to give consistent results.

Several key points have emerged from these early studies. First, notwithstanding the uncertainties, both the lensing and HDF photometric redshift data point to a low mean redshift (z$\approx$1-2) for the I$\approx$26 population. The addition of near-infrared photometry to the HDF data appears to make the agreement with the lensing results even more convincing (Connolly et al, in preparation). Second, there is the small physical size (2--4 h$^{-1}$ kpc) of the faint star-forming population. Given that the bulk of the integrated light in the number counts arises from brighter systems with z<1 (Section 3.1), the faint population beyond the number count break appears to represent an era of initial star formation at modest redshift z<2. The complex morphology of many of the higher z luminous systems (Abraham et al 1996a, Giavilisco et al 1996), and the large ages inferred for spheroidal populations, at least in clusters (Bower et al 1992, Ellis et al 1996b), supports the suggestion that bulges form early and disk galaxies assemble by gradual infall and accretion (Cowie 1988, Baugh et al 1996). Specific cases where galaxies may be assembling in this way at high redshift have been proposed by Pascarelle et al (1996).

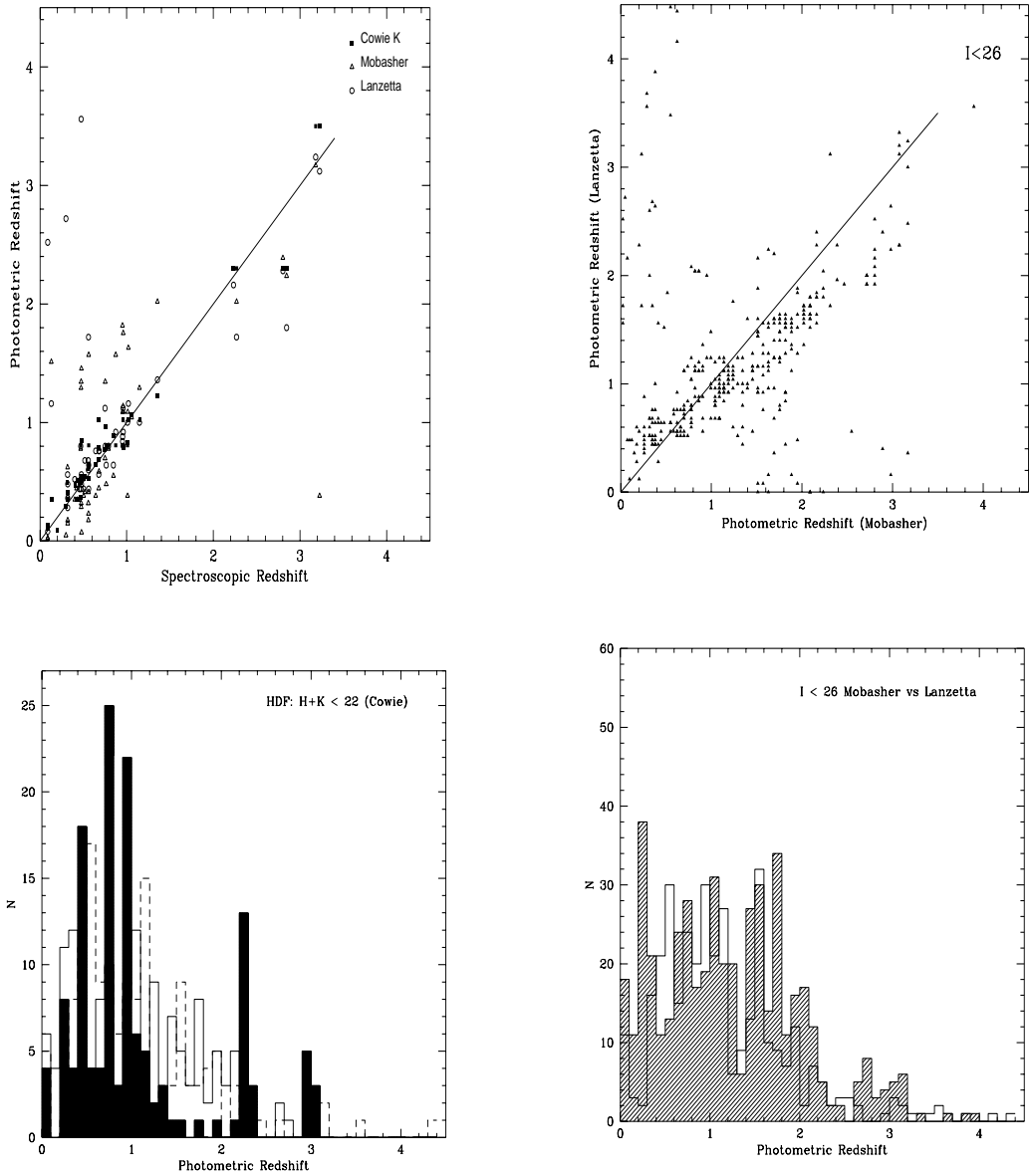

*Figure 9: A study of various photometric redshift catalogs available for the HDF (see Table 2): (a) photometric redshifts for 3 groups compared with currently available Keck spectroscopic values. Note that Cowie's (1996) estimates take advantage of deep infrared photometry. (b) Distribution of photometric redshifts to a limit of H+K<22 (Cowie: filled, Lanzetta et al (1996): solid line, Mobasher et al (1996): dashed line). (c) A comparison of photometric redshifts to I<26. (d) Distribution of photometric redshifts to I<26 (Mobasher et al: shaded, Lanzetta et al: solid line).*

As with the z<1 data however, to make the connection with local systems, some reliable stable physical property such as mass is required. In fact, little is presently known about the dynamics of the most distant systems. Attempts to estimate masses from absorption line profiles are limited by a poor understanding of whether line broadening is due to shocks or turbulent motions. Examination of gravitationally-magnified samples (c.f. Yee et al 1996, Ebbels et al 1996, Williams & Lewis 1996) will be particularly profitable as the boost of 2-3 mag will enable very detailed line studies.

## 6. THE STAR FORMATION HISTORY OF FIELD GALAXIES

In the past year, there has been a remarkable synthesis of the star formation history of field galaxies (as delineated by the observations reviewed here) and the gaseous and chemical evolution of the intergalactic gas (as delineated by the studies of QSO absorption lines). Lanzetta et al (1995), Wolfe et al (1995) and more recently Storrie-Lombardi et al (1996) analyzed the evolution of the cosmic density of neutral hydrogen with redshift via various QSO absorber samples. These data locate a redshift of z=2-3 where the bulk of the present star formation density is seen in neutral hydrogen clouds. Likewise, the chemical evolution of metallic clouds has been studied with redshift by Pettini et al (1994). Pei & Fall (1995) and Fall et al (1996) have shown, via a remarkably simple model, how these various data can be reconciled with the history of the volume-averaged star formation. Madau et al (1996) have updated this analysis with the most recent estimates of the high redshift SFR discussed in Section 5 (Figure 10). The emerging picture points to a redshift range of 1-2 in which the bulk of the present-day stellar population was assembled and perhaps most of the present-day metals produced (Cowie 1988, Songaila et al 1990, Ellis 1996b).

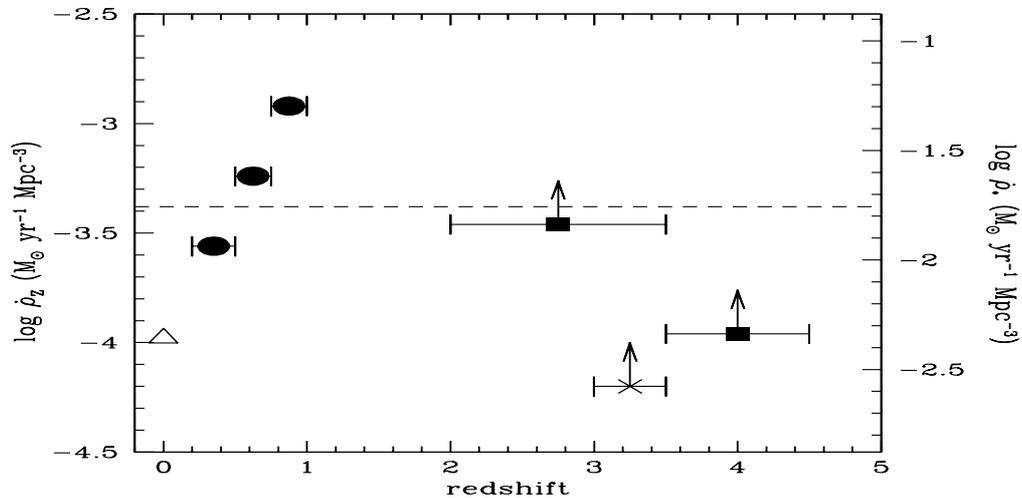

*Figure 10: Element and star formation history from the analysis of Madau et al (1996). Data points provide a measurement or a lower limit to the universal metal ejection rate (left ordinate) and total star formation density (right ordinate). Triangle: local Hα survey of Gallego et al (1995). Filled dots: CFRS redshift survey Lilly et al (1996). Diagonal cross: lower limit from Lyman limit galaxies observed by Steidel et al (1996b). Filled squares: similar lower limts from HDF data available in mid-1996. The dashed line depicts the fiducial rate equivalent to the mass density of local metals divided by the present age of the Universe. ($\Omega=1$, $H_o = 50$ kms$^{-1}$ Mpc$^{-1}$)*

Appealing as this picture is, not least because it has much theoretical support (Kauffmann et al 1996, Baugh et al 1996, White 1996), it rests on preliminary observations and relies on the connection of disparate datasets, each using a different indicator of star formation and each selecting only some detectable subset of the overall population. At low z, is the star formation density derived from the Hα surveys (Gallego et al 1995) affected by fair sample problems that have been invoked to address the apparent low normalisation of the local LF (Section 4.1)? At modest redshift, the rapid increase in the integrated star formation density (Lilly et al 1996, Cowie et al 1996) relies on the conversion of blue light or [O II] line strength to the SFR and, quantitatively, via the extrapolation of the contribution to sources whose luminosities are fainter than the redshift survey limit. These uncertainties may affect the rate of increase with redshift beyond z=0.8 (Lilly et al 1996). It will therefore be important to extend the redshift surveys fainter even at z<1 to verify the extrapolation as well as to use alternative diagnostics at moderate redshift such as Hα fluxes secured via infrared spectroscopy. There are also promising new approaches to constrain the intermediate z SFR based entirely on emission line searching (Meisenheimer et al 1996). Similarly, at high z, the LF and spatial distribution of Lyman limit-selected samples needs to be explored via panoramic surveys based on the highly successful pilot studies in the HDF (Steidel et al 1996b). The star formation densities derived from these data must be regarded as lower limits until the effects of dust have been properly explored.

Despite these caveats, the trend is highly encouraging. A clear gap emerges, however, in our knowledge in the redshift range between that reviewed here at z<1 and that revealed at z>2.3 in the HST work. This has motivated the construction of a new generation of ground-based infrared spectrographs free from OH background light (Iwamuro et al 1994, LeFevre et al 1996c, Taylor & Colless 1996, Piche et al 1997,) which aim to survey this difficult region in conjunction with optical-UV imaging. Together with NICMOS on HST which should have powerful background-limited capabilities in both deep imaging and low resolution grism spectroscopic modes, the intermediate redshift range can soon be systematically explored in a way that was highly successful for the z<1 population.

The non-zero metallicity of the highest redshift absorbing clouds (Cowie 1996) also points to an earlier era of modest star formation possibly associated with the small but convincing population of high z galaxies already known to contain old stars (Dunlop et al 1996, Stockton et al 1995). The detection of high redshift sources at far infrared and sub-millimeter wavelengths (Omont et al 1996) together with the successful deployment of the Infrared Space Observatory and the SCUBA sub-millimeter array detector (Gear & Cunningham 1995) augurs well for surveying the z>4 universe for earlier eras of star formation. This will remain an important observational challenge even if the bulk of the SF activity is convincingly demonstrated to occur at z=1-2. To physically understand the processes that lead to galaxy formation, it will surely be necessary to explore the high z tail.

Tremendous progress has been made in observational cosmology. The subject of galaxy formation and evolution has moved firmly from the realm of theoretical speculation into that of systematic observation. Some lessons are, however, being learnt. Perhaps the most important of these is the dangers of relying purely on morphology and star formation diagnostics to connect what may, in fact, be very different populations observed via different techniques at high and low z. Clearly we seek more representative physical parameters to sub-classify the datasets over a range of look-back times in order to test detailed physical hypotheses. A further hindrance is the absence of well-defined local data of the kind needed for detailed comparisons with the high z samples (Koo & Kron 1992). However, notwithstanding the formidable challenges of studying the distant universe, the combination of deeper redshift surveys and morphologies from HST has demonstrated quantitatively the presence of rapid evolution in a subset of the population to z=1. The absence of a dominant population of star forming galaxies at z=3 and the small physical sizes of the faint HDF images delineates a simple picture that is consistent with hierarchical galaxy formation and knowledge of the properties of intervening gas clouds as studied in QSO absorption lines. Galactic history seems to have been remarkably recent which can only be our good fortune given the power of our new facilities to observe these eras in considerable detail. The *observational picture* is already emerging very rapidly but much work and ingenuity will be needed to identify the *physical processes* that drive the evolutionary trends now revealed.

ACKNOWLEDGEMENTS

I thank my collaborators, colleagues and visitors at Cambridge, particularly Roberto Abraham, Matthew Colless, George Efstathiou, Masataka Fukugita, Simon Lilly and Max Pettini for their critical and helpful comments on this review. Many workers sent detailed accounts of their views on the sensitive and complex issues discussed in Section 4. Special thanks are due to Emmanuel Bertin, Ray Carlberg, Stephane Charlot, Len Cowie, Julianne Dalcanton, Simon Driver, Harry Ferguson, Luigi Guzzo, Karl Glazebrook, David Hartwick, David Koo, Huan Lin, Stacy McGaugh, Nigel Metcalfe, John Peacock, Bianca Poggianti, Tom Shanks and Elena Zucca. I apologise to these and others if I have failed to represent their particular viewpoint fairly. I thank Bernard Sadoulet, Joe Silk and Tom Broadhurst for their hospitality and support in Berkeley during the summer of 1996 when much of this review was written. Finally I thank Allan Sandage for his numerous suggestions and encouragement.